\documentclass[twoside,twocolumn,9pt]{article}
\usepackage{extsizes}
\usepackage[super,sort&compress,comma]{natbib} 
\usepackage[version=3]{mhchem}
\usepackage[left=1.5cm, right=1.5cm, top=1.785cm, bottom=2.0cm]{geometry}
\usepackage{balance}
\usepackage{mathptmx}
\usepackage{sectsty}
\usepackage{graphicx} 
\usepackage{lastpage}
\usepackage[format=plain,justification=justified,singlelinecheck=false,font={stretch=1.125,small,sf},labelfont=bf,labelsep=space]{caption}
\usepackage{float}
\usepackage{fancyhdr}
\usepackage{fnpos}
\usepackage[english]{babel}
\addto{\captionsenglish}{%
  
}
\usepackage{array}
\usepackage{droidsans}
\usepackage{charter}
\usepackage[T1]{fontenc}
\usepackage[usenames,dvipsnames]{xcolor}
\usepackage{setspace}
\usepackage[compact]{titlesec}
\usepackage{hyperref}
\usepackage{subcaption}
\newcommand{\subfiglabel}[1]{\begin{subfigure}{0em}\phantomsubcaption\label{#1}\end{subfigure}}
\usepackage{siunitx} 
\DeclareSIUnit\bar{bar} 

\definecolor{cream}{RGB}{222,217,201}

\begin{document}

\pagestyle{fancy}
\thispagestyle{plain}
\fancypagestyle{plain}{
\renewcommand{\headrulewidth}{0pt}
}

\makeFNbottom
\makeatletter
\renewcommand\LARGE{\@setfontsize\LARGE{15pt}{17}}
\renewcommand\Large{\@setfontsize\Large{12pt}{14}}
\renewcommand\large{\@setfontsize\large{10pt}{12}}
\renewcommand\footnotesize{\@setfontsize\footnotesize{7pt}{10}}
\makeatother

\renewcommand{\thefootnote}{\fnsymbol{footnote}}
\renewcommand\footnoterule{\vspace*{1pt}%
\color{cream}\hrule width 3.5in height 0.4pt \color{black}\vspace*{5pt}} 
\setcounter{secnumdepth}{5}

\makeatletter 
\renewcommand\@biblabel[1]{#1}            
\renewcommand\@makefntext[1]%
{\noindent\makebox[0pt][r]{\@thefnmark\,}#1}
\makeatother 
\renewcommand{\figurename}{\small{Fig.}~}
\sectionfont{\sffamily\Large}
\subsectionfont{\normalsize}
\subsubsectionfont{\bf}
\setstretch{1.125} 
\setlength{\skip\footins}{0.8cm}
\setlength{\footnotesep}{0.25cm}
\setlength{\jot}{10pt}
\titlespacing*{\section}{0pt}{4pt}{4pt}
\titlespacing*{\subsection}{0pt}{15pt}{1pt}

\fancyfoot{}
\fancyfoot[LO,RE]{\vspace{-7.1pt}\includegraphics[height=9pt]{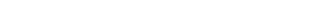}}
\fancyfoot[CO]{\vspace{-7.1pt}\hspace{11.9cm}\includegraphics{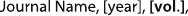}}
\fancyfoot[CE]{\vspace{-7.2pt}\hspace{-13.2cm}\includegraphics{head_foot/RF}}
\fancyfoot[RO]{\footnotesize{\sffamily{1--\pageref{LastPage} ~\textbar  \hspace{2pt}\thepage}}}
\fancyfoot[LE]{\footnotesize{\sffamily{\thepage~\textbar\hspace{4.65cm} 1--\pageref{LastPage}}}}
\fancyhead{}
\renewcommand{\headrulewidth}{0pt} 
\renewcommand{\footrulewidth}{0pt}
\setlength{\arrayrulewidth}{1pt}
\setlength{\columnsep}{6.5mm}
\setlength\bibsep{1pt}

\makeatletter 
\newlength{\figrulesep} 
\setlength{\figrulesep}{0.5\textfloatsep} 

\newcommand{\topfigrule}{\vspace*{-1pt}%
\noindent{\color{cream}\rule[-\figrulesep]{\columnwidth}{1.5pt}} }

\newcommand{\botfigrule}{\vspace*{-2pt}%
\noindent{\color{cream}\rule[\figrulesep]{\columnwidth}{1.5pt}} }

\newcommand{\dblfigrule}{\vspace*{-1pt}%
\noindent{\color{cream}\rule[-\figrulesep]{\textwidth}{1.5pt}} }

\makeatother

\twocolumn[
  \begin{@twocolumnfalse}
{\includegraphics[height=30pt]{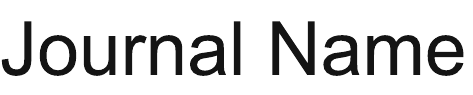}\hfill\raisebox{0pt}[0pt][0pt]{\includegraphics[height=55pt]{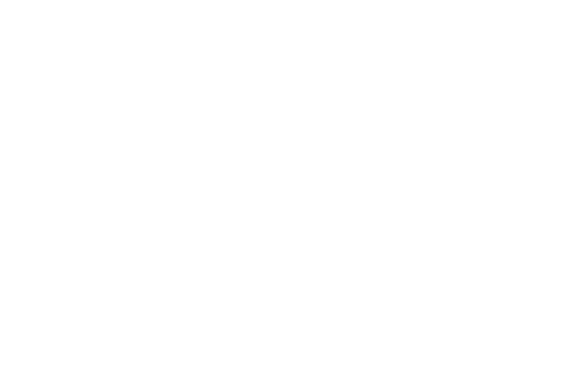}}\\[1ex]
\includegraphics[width=18.5cm]{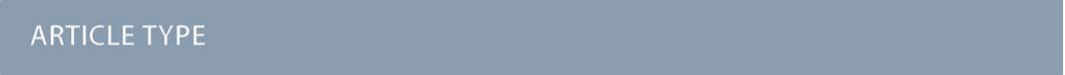}}\par
\vspace{1em}
\sffamily
\begin{tabular}{m{4.5cm} p{13.5cm} }

\includegraphics{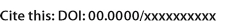} & \noindent\LARGE{\textbf{Silicon oxide nanoparticles grown on graphite by co-deposition of the atomic constituents$^\dag$}} \\
\vspace{0.3cm} & \vspace{0.3cm} \\

 & \noindent\large{Steffen Friis Holleufer,$^{\ast}$\textit{$^{a,b}$} Alfred Hopkinson,\textit{$^{a}$} Duncan S. Sutherland,\textit{$^{b}$} Zheshen Li,\textit{$^{c}$} Jeppe V. Lauritsen,\textit{$^{b}$} Liv Hornekær,\textit{$^{a,b}$} and Andrew Cassidy\textit{$^{a}$}} \\

\includegraphics{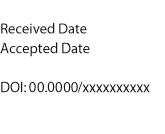} & \noindent\normalsize{Nanoscale silicate dust particles are the most abundant refractory component observed in the interstellar medium and thought to play a key role in catalysing the formation of complex organic molecules in the star forming regions of space. 
We present a method to synthesise a laboratory analogue of nanoscale silicate dust particles on highly oriented pyrolytic graphite (HOPG) substrates by co-deposition of the atomic constituents. 
The resulting nanoparticulate films are sufficiently thin and conducting to allow for surface science investigations, and are characterised here, \textit{in situ} under UHV, using X-ray photoelectron spectroscopy, near-edge X-ray absorption atomic fine spectroscopy and scanning tunnelling microscopy, and, \textit{ex situ}, using scanning electron microscopy. 
We compare \ce{SiO_x} film growth with and without the use of atomic O beams during synthesis and conclude that exposure of the sample to atomic O leads to homogeneous films of interconnected nanoparticle networks. 
The networks covers the graphite substrate and demonstrate superior thermal stability, up to \qty{1073}{\kelvin}, when compared to oxides produced without exposure to atomic O. 
In addition, control over the flux of atomic O during growth allows for control of the average oxidation state of the film produced. 
Photoelectron spectroscopy measurements demonstrate that fully oxidised films have an \ce{SiO2} stoichiometry very close to bulk \ce{SiO2} and scanning tunnelling microscopy images show the basic cluster building unit to have a radius of approximately \qty{2.5}{\nm}. 
The synthesis of \ce{SiO_x} films with adjustable stoichiometry and suitable for surface science experiments that require conducting substrates will be of great interest to the astrochemistry community, and will allow for nanoscale-investigation of the chemical processes thought to be catalysed at the surface of dust grains in space.} \\

\end{tabular}

 \end{@twocolumnfalse} \vspace{0.6cm}

  ]

\renewcommand*\rmdefault{bch}\normalfont\upshape
\rmfamily
\section*{}
\vspace{-1cm}


\footnotetext{\textit{$^{a}$~Center for Interstellar Catalysis, Department of Physics and Astronomy, Aarhus University, 8000 Aarhus C, Denmark.}}
\footnotetext{\textit{$^{b}$~Interdisciplinary Nanoscience Center, Aarhus University, 8000 Aarhus C, Denmark.}}
\footnotetext{\textit{$^{c}$~Centre for Storage Ring Facilities, Department of Physics and Astronomy, Aarhus University, 8000 Aarhus C, Denmark.}}
\footnotetext{\dag~Supplementary Information available: XPS data for C1s core levels and large scale SEM images of silicon oxide-covered HOPG are available as supplementary information.}


\section{Introduction}
Across the Interstellar Medium (ISM), Si and O are observed to be heavily depleted from the gas phase and are proposed to condense and form refractory silicate particles.\cite{vansteenberg1988,tielens1998,draine2003,henning2010,zhukovska2016,zhukovska2018}
These silicate particles are typically thought to have dimensions on the scale of hundreds of nanometres\cite{draine2003,demyk2011} and are commonly referred to as interstellar dust.
Interstellar silicate dust materials play essential roles in the physical and chemical processes that occur in star forming regions of the ISM, including catalysing the formation of molecular hydrogen\cite{hollenbach1971,vidali2009,vidali2013,suhasaria2021} and water\cite{serraperalta2022}, and acting as substrates for the growth of molecular ices.\cite{mcclure2024}
While the chemical and physical properties of interstellar silicates differ depending on their environment and location in the interstellar life cycle of matter, they are generally considered as amorphous \ce{SiO_x} or \ce{SiO2} lattices rich in Mg and doped with Fe.\cite{henning2010,voshchinnikov2010}
Formation of \ce{SiO_x} is expected to proceed by gas-phase formation of \ce{SiO} molecules,\cite{he2022} which subsequently condense to form \ce{SiO_x} clusters and particles with a variety of stoichiometries.\cite{reber2008,henning2017}
Pure \ce{SiO2} particles are not detected in the ISM in significant amounts,\cite{li2002} but have been observed in protoplanetary disks.\cite{sargent2009}
It has recently been suggested that silicate dust particles might be mixed with other interstellar dust species, such as carbon-rich dust. 
These mixed-dust structures are referred to as "astrodust".\cite{Draine2021,Hensley2023}

Despite their importance to astrochemistry, there is a lack of laboratory model systems of silicate dust materials that would allow for experimental surface science investigations into their catalytic properties. 
Bulk silicate materials are electrical insulators with band gaps above \qty{8.0}{\electronvolt},\cite{schneider1976} hindering their use in scanning tunnelling microscopy (STM) and photoelectron spectroscopy experiments. 
An ideal laboratory analogue of silicate dust would i) facilitate studies of both isolated silicate material and silicate material mixed with carbon, representing mixed astrodust; ii) give control over the oxidation state of the silicon component, making it possible to synthesise silicate dust materials with stoichiometries between \ce{SiO_x} and \ce{SiO2}; iii) facilitate incorporation of Mg and Fe cations, to test the catalytic potential of these materials and test their significance in interstellar silicate dust chemistry; and iv) be sufficiently conducting so as to allow for traditional surface science investigations that would give an atomic understanding of chemical and physical processes involving dust.
With this paper, we introduce a laboratory analogue for silicate dust, synthesised using the atomic components, namely silicon and oxygen, on a graphitic substrate. 

Terrestrially, silicon oxides and minerals have high natural abundance\cite{shaw1967,shaw1976,mcdonough1995,wedepohl1995} and are of technological interest in many fields including energy storage, electronics, and catalysis. 
Silicon dioxide (\ce{SiO2}), or silica, is the stoichiometry obtained for fully oxidised Si.
Silica exists in a variety of crystalline or amorphous structures, that all contain the same basic building block, \textit{i.e.}, corner-sharing \ce{[SiO4]^4-} tetrahedra. 
Even when Si atoms are not fully oxidised, resulting in Si-suboxide structures (\ce{SiO_x} with $x < 2$), Si atoms remain tetrahedrally coordinated. 
Herein, \ce{Si-Si} bonds can be formed, in contrast to the fully oxidised \ce{SiO2} structure.\cite{shallenberger1996} 
The number of \ce{Si-O} bonds formed on average by an individual silicon atom in an \ce{SiO_x} material is commonly referred to as its oxidation state. 
Si(0) denotes Si bound only to other Si-atoms, while Si(IV) represents Si bound to four O-atoms, as is the case for \ce{SiO2}.
The intermediate states Si(I), Si(II), and Si(III) then constitute a Si-atom bound to one, two, and three O-atoms, respectively. 
The remaining $4 - n$ bonds of Si($n$) are \ce{Si-Si} bonds.

Bulk silicon oxides are commonly investigated as high surface area support materials for heterogeneous catalysts of importance in numerous industrial reactions,\cite{verma2020} \textit{e.g.}, the Fischer--Tropsch process\cite{khodakov2002,liu2024}, hydrodeoxygenation,\cite{perezestrada2024} and nitrile hydrogenation.\cite{chandrashekhar2022}
A thorough understanding of processes like metal doping, molecular reactions, and structural decomposition of silicon oxides require the development of model systems with adequate structural tunability.
2D silica films are examples of silicon oxide model systems, and these 2D substrates have attracted considerable interest in recent years.
2D silicas can be readily grown on various metallic substrates, like Mo(112)\cite{weissenrieder2005}, Ru(0001),\cite{loffler2010,yang2012} Pd(100),\cite{altman2013} Pt(111),\cite{yu2012,crampton2015} Au(111),\cite{doudin2022} and others.\cite{jhang2017,tissot2018,krinninger2024}
The resulting 2D silica film can be considered as either a free-standing bilayer or a substrate-supported monolayer of \ce{[SiO4]^4-} tetrahedra, with tunability offered by choice of substrate.
The silica films can be further tuned to be either amorphous or crystalline and can be doped with metals, \textit{e.g.}, Fe,\cite{wlodarczyk2013} effectively making a 2D silicate.
Importantly, these 2D silicas are sufficiently thin to be characterised using surface science tools that require an electrically conducting substrate and thus allow for atomic-level studies of their structure. For a thorough discussion on 2D silicas, the reader is referred to one of the various reviews and perspectives on the topic, as well as references therein.\cite{buchner2017,zhong2022,altman2024}

2D silicas provide a unique opportunity to study silicon oxides as support materials for metal atoms and NPs.
However, none of the existing substrates proposed for supporting the growth of 2D silicas are suitable for use as laboratory analogues for Interstellar dust. 
For the study of chemical reactions on silica films, concerns might arise regarding the catalytic activity of the underlying metallic substrates, likely making the model system irrelevant for astrochemical studies.
This is particularly important for films grown on Pt, Pd, or Ru, where the substrate has a high catalytic activity.
Thus, there is considerable interest in synthesising a similar silicon oxide model system on an inert substrate of more relevance to astrochemistry, such as graphite.
The inherent chemical inertness of highly oriented pyrolytic graphite (HOPG) and synthesis of a \ce{SiO_x}/HOPG system would provide a laboratory analogue of silicate dust materials and could act as a model system for astrodust studies. 
A 2D silica on graphene supported on Cu-foil has previously been reported,\cite{huang2012} and this report prompted us to attempt to synthesise a thin silica film on HOPG.

In the following we demonstrate the growth of a silicon oxide model system on HOPG by co-depositing atomic Si and atomic O, and show that this strategy produces a laboratory analogue of interstellar dust that meets the criteria defined above. 
We demonstrate a highly oxidised, nanostructured \ce{SiO_x} system ($x\approx2$) that covers the HOPG surface and is thermally stable up to \qty{1073}{\kelvin}. 
\ce{SiO_x} films produced with the use of just molecular oxygen or with molecular and atomic oxygen combined are compared, and the inclusion of atomic oxygen during growth is shown to produce more stable films and allow for control over the average oxidation state of the Si constituent via adjustment of the relative Si-atom / O-atom flux during growth.
The synthesised materials are characterised using X-ray photoelectron spectroscopy (XPS) and the nanoscale structure and morphology are characterised using STM and scanning electron microscopy (SEM). 

\section{Experimental Methods}
\subsection{Preparation of SiO$_\text{x}$/HOPG}
Experiments were conducted in UHV chambers operating at base pressures below \qty{1E-09}{\milli\bar}.
Clean HOPG samples (Goodfellow Cambridge Ltd) were prepared by cleaving in air followed by annealing above \qty{1100}{\kelvin} in UHV using e-beam annealing.
The cleanliness of the HOPG surface was confirmed with XPS.
Temperatures were measured at the back of the HOPG substrate with a K-type thermocouple.

Growth of \ce{SiO_x} on HOPG proceeded by co-depositing beams of Si and O.
A beam of Si atoms and ions was produced by e-beam evaporation of a Si rod ($\geq99.999\%$, Goodfellow Cambridge Ltd) using a commercial e-beam evaporator (EFM 3T, FOCUS GmbH).
The flux of Si was monitored with the built-in flux monitor of the e-beam evaporator.
The deposition rate of Si on HOPG was evaluated in \qty{}{\angstrom\per\second} by depositing Si on clean HOPG.
The resulting film thickness was calculated by modelling the attenuation of the HOPG C~1s core-level spectrum, ${\frac{I}{I_0}}$, assuming a homogeneous Si film of thickness $t$, using the equation $t = -\lambda(E) \ln{\frac{I}{I_0}}$.
Here, $\lambda(E)$ is the inelastic mean free path (IMFP) through Si of photoelectrons with energy $E$, excited from the C~1s core-level.
The thicknesses of \ce{SiO2} films were similarly calculated using the IMFP of photoelectrons through \ce{SiO2}.
IMFPs were obtained from the NIST electron inelastic-mean-free-path database.\cite{jablonski2010,tanuma1991}
The deposition rate density of Si in \qty{}{atoms \per\centi\meter\squared\per\second} was estimated via conversion from the measured growth rate in \qty{}{\angstrom\per\second}.
The conversion accounts for the atom density and thickness of a Si monolayer.
Monolayers with facets of Si(100) ($t_\text{ML} = $ \qty{1.36}{\angstrom}, \qty{6.78E14}{atoms \per\centi\meter\squared}), Si(110) ($t_\text{ML} = $ \qty{1.92}{\angstrom}, \qty{9.59E14}{atoms \per\centi\meter\squared}), and Si(111) ($t_\text{ML} = $ \qty{1.57}{\angstrom}, \qty{7.83E14}{atoms \per\centi\meter\squared}) were considered.
The choice of facet did not have a notable impact on the resulting deposition rate density.

The O-atom beam was produced by a commercial oxygen atom beam source (OABS, MBE Komponenten GmbH) via thermal cracking of \ce{O2} ($\geq99.9\%$) passing through a heated Ir capillary.
The O-atom flux in \qty{}{atoms\per\centi\meter\squared\per\second} was estimated from the OABS operational parameters (\ce{O2} feeding pressure, Ir capillary temperature and distance to sample), as discussed for H-atom beam generation in \citeauthor{tschersich1998}\cite{tschersich1998} and \citeauthor{tschersich2000}.\cite{tschersich2000}
The relative flux of atomic O versus molecular O at the sample was controlled by changing the temperature of the Ir capillary to adjust the cracking efficiency, while maintaining the same feeding pressure of \ce{O2}.

Overlapping deposition areas of Si and O were confirmed by aligning the sample with the overlapping lights from the e-beam evaporator filament and a torch shining through the OABS.

\subsection{Sample characterisation}
\subsubsection*{X-ray photoelectron spectroscopy.~~}
Laboratory XPS measurements were taken using a non-monochromatic Al K$\alpha$ (photon energy \qty{1486.6}{\electronvolt}) source (XR 50, SPECS GmbH) and a hemispherical electron energy analyser (PHOIBOS 100 MCD5, SPECS GmbH).
The angle between X-ray source and analyser was \qty{54.5}{\degree}.
All spectra were captured at a take-off angle of \qty{90}{\degree} in the High Magnification lens mode.
The detector voltage was set at \qty{2550}{\volt}, and a pass energy of \qty{20}{\electronvolt} was used.
O~1s and Si~2p spectra were captured with a step size of \qty{0.07}{\electronvolt}.
C~1s spectra were captured with a step size of \qty{0.05}{\electronvolt}.

Synchrotron radiation XPS (SR-XPS) measurements were performed at the AU-Matline end-station of the ASTRID2 synchrotron (Aarhus University, Aarhus, Denmark) using a hemispherical electron energy analyser (PHOIBOS 150 1D-DLD, SPECS GmbH).
The angle between X-ray beam and analyser was \qty{45}{\degree}.
SR-XPS spectra were recorded at normal emission.
All SR-XPS spectra have been normalised to the beam current of the synchrotron storage ring.

XPS data was fitted using the KolXPD software.\cite{kolxpd}
Visualisation of data and calculation of relative peak areas was performed using Python~3 (NumPy and Matplotlib).

\subsubsection*{Near-edge X-ray absorption fine structure.~~}
NEXAFS measurements were performed at the AU-Matline end-station of the ASTRID2 synchrotron (Aarhus University, Aarhus, Denmark) using a hemispherical electron energy analyser (PHOIBOS 150 1D-DLD, SPECS GmbH).
All spectra presented were captured with total electron yield (TEY) measurements with electron emission normal to the substrate surface.

\subsubsection*{Scanning electron microscopy.~~}
For SEM images, the sample was transferred from the UHV chamber to the cleanroom at the Interdisciplinary Nanoscience Center (iNANO, Aarhus University). Secondary electron images were captured using a Magellan 400 Field Emission Scanning Electron Microscope (FEI Company) using a \qty{5}{\kilo\electronvolt} beam with nominal beam current of \qty{50}{\pico\ampere} and have not been processed further.

\subsubsection*{Scanning tunnelling microscopy.~~}
STM images were captured using a home-built Aarhus-type STM located in the same UHV chamber as the laboratory XPS setup. 
Data were flattened prior to analysis using Gwyddion.\cite{gwyddion}

\begin{figure*}[ht]
    \centering
    \includegraphics[]{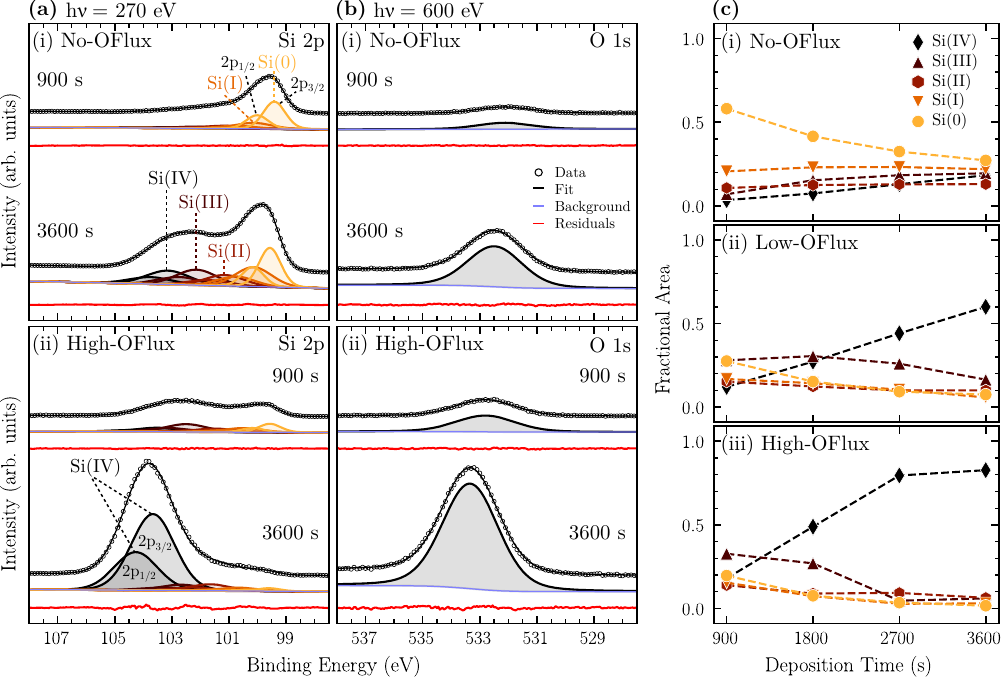}
    \subfiglabel{fgr:O_vs_O2_Si2p}
    \subfiglabel{fgr:O_vs_O2_O1s}
    \subfiglabel{fgr:O_vs_O2_Si2pArea}
    \caption{SR-XPS spectra of the (a) Si~2p and (b) O~1s core-levels, following the growth of films of \ce{SiO_x} on a HOPG substrate in absence, (i) - No-OFlux; and presence (ii) - High-OFlux of atomic oxygen. In both cases spectra were recorded after \qty{900}{\second} and \qty{3600}{\second} co-deposition of Si and oxygen on HOPG, see the main text for details. Backgrounds for each spectrum are plotted as blue lines. Data are shown as black circles and have been fitted with curves representing Gaussian-Lorentzian convolutions. Fitting accounts for spin-orbit splitting of Si~2p with $\Delta =$ \qty{0.6}{\electronvolt} and a 1:2 area ratio for Si~2p$_{1/2}$ and 2p$_{3/2}$. Fitting residuals are shown below each spectrum.
    Data have been offset in intensity and the number of plotted data points has been reduced by a factor of 2 for clarity. 
    The Y-axes have been scaled for all spectra to allow relative comparisons between samples by assuming an equal amount of C~1s intensity for clean HOPG.
    Residuals are plotted on the same scale.
    (c) The evolution of the fractional peak area for each component in fits to the photoelectron spectra from the Si~2p core level are shown as a function of deposition time for samples (i) No-OFlux, (ii) Low-OFlux, and (iii) High-OFlux.}
    \label{fgr:O_vs_O2}
\end{figure*}

\section{Results and Discussion}
\subsection{Growth of SiO$_\text{x}$ films} \label{subsect:SiOxGrowth}
To quantify the impact of the O-atom beam flux on the growth of \ce{SiO_x} films, three \ce{SiO_x}/HOPG samples were prepared with different ratios between the atomic and molecular oxygen fluxes used during growth.
The Si deposition rate, estimated to be \qty{1E12}{atoms \per\centi\meter\squared\per\second}, and the \ce{O2} feeding pressure provided to the OABS, \qty{1.2E-02}{\milli\bar}, resulting in chamber pressures between \qtyrange[range-units=single]{2E-07}{3E-07}{\milli\bar}, were kept constant for all three samples.
The efficiency of \ce{O2} cracking to form atomic O was regulated by operating the OABS at RT, \qty{1843}{\kelvin} and \qty{1943}{\kelvin}.
The resulting three samples are referred to as No-OFlux (T$_\text{OABS} =$ \qty{300}{\kelvin}, 0 \% \ce{O2} cracking, \qty{0}{O-atoms \per\centi\meter\squared\per\second}), Low-OFlux (T$_\text{OABS} =$ \qty{1843}{\kelvin}, 2.1 \% \ce{O2} cracking, \qty{8E12}{O-atoms \per\centi\meter\squared\per\second}), and High-OFlux (T$_\text{OABS} =$ \qty{1943}{\kelvin}, 4.8 \% \ce{O2} cracking, \qty{2E13}{O-atoms \per\centi\meter\squared\per\second}).

XPS measurements for the three samples are presented in Figure~\ref{fgr:O_vs_O2}.
Spectra were recorded as a function of fluence, following co-deposition times of \qty{900}{\second}, \qty{1800}{\second}, \qty{2700}{\second}, and \qty{3600}{\second}.
Figure~\ref{fgr:O_vs_O2_Si2p} shows the photoelectron count at the Si~2p core-level for samples No-OFlux (Figure~\ref{fgr:O_vs_O2_Si2p}(i)) and High-OFlux (Figure~\ref{fgr:O_vs_O2_Si2p}(ii)).
The spectra were captured with a photon energy of \qty{270}{\electronvolt}.
After \qty{900}{\second} of Si deposition in \ce{O2} for sample No-OFlux, the Si~2p core-level signal exhibits a peak with binding energies between \qtyrange[range-units=single]{99.5}{100}{\electronvolt}, corresponding to elemental Si, Si(0).
Small amounts of Si oxidation are observed as a tail toward higher binding energy.
After \qty{3600}{\second} of Si deposition in \ce{O2}, the binding energy of the Si~2p core-level remains between \qtyrange[range-units=single]{99.5}{100}{\electronvolt}, with a broad oxide feature observed at higher binding energy, with a maximum at \textit{circa} \qty{102}{\electronvolt}.
The signal was fitted using a model that accommodated five Si-states, \textit{i.e.}, from elemental Si, Si(0), to fully oxidised Si, Si(IV), with the resulting fit and fit components indicated in Figure~\ref{fgr:O_vs_O2_Si2p}(i).
This model suggests that the film contains a mixture of all Si oxidation states, with Si(0) as the dominant component.

In contrast, the Si~2p core-level signal from the sample prepared with the highest fluence of atomic O, labelled High-OFlux, exhibits a two-peak structure after co-deposition for \qty{900}{\second}, indicating the presence of both elemental Si and \ce{SiO_x} species.
After \qty{3600}{\second}, the Si~2p core-level signal exhibits a clear peak close to \qty{104}{\electronvolt}, with a small shoulder between \qtyrange[range-units=single]{100}{102}{\electronvolt}. 
Fitting this dataset with the same peak-fitting model indicates that Si(IV) is the dominant component, while Si(0) and the suboxidic species are present in very small amounts.
The final film thickness was calculated to be \qtyrange[range-units=single]{1.5}{2}{\angstrom} using the attenuation of the HOPG C~1s core-level spectrum.

Spectra, from the No-OFlux and High-OFlux samples, showing the photoelectron flux from the O~1s core-level, obtained using a photon energy of \qty{600}{\electronvolt}, are presented in Figure~\ref{fgr:O_vs_O2_O1s}. 
When co-depositing Si and \ce{O_2}, in the absence of atomic O, the O~1s core-level spectrum exhibits a small peak at \qty{532.1}{\electronvolt} after \qty{900}{\second}, which intensifies and shifts to \qty{532.5}{\electronvolt} after \qty{3600}{\second}, as shown in Figure~\ref{fgr:O_vs_O2_O1s}(i).
The signal can be fitted with a single \ce{O-Si} component with a FWHM of \qty{2.25}{\electronvolt}.
Figure~\ref{fgr:O_vs_O2_O1s}(ii) shows the O~1s core-level spectrum for sample High-OFlux.
After \qty{900}{\second}, the core-level signal can be fitted with a single \ce{O-Si} component with a maximum at \qty{532.8}{\electronvolt} and a FWHM of \qty{2.45}{\electronvolt}, and this peak shifts to \qty{533.3}{\electronvolt} with a FWHM of \qty{2.30}{\electronvolt} after \qty{3600}{\second}. 
We note that neither sample shows any indication of chemical groups associated with functionalisation of the HOPG, \textit{e.g.} an epoxy-group component at \qty{286.2}{\electronvolt}.\cite{barinov2009,vinogradov2011,larciprete2012} 
Instead, the C~1s core-level spectra measured for all samples, including Low-OFlux, are strikingly similar, suggesting no bonding between C and O, Figure~S1. 
The broader O~1s peak observed for film growth with atomic O is attributed to a larger variety of chemically similar, but slightly distinct, O-atoms in the resulting \ce{SiO_x} film.
The relative intensity between the O~1s core-level signal and the Si~2p signal is noticeably higher for High-OFlux sample compared to the No-OFlux sample. 
This confirms the higher degree of oxidation of the Si caused by co-deposition with atomic O, in agreement with the analysis of the Si~2p core-level spectra. 
The observed shifts in the O~1s core-level spectra with increasing fluence are likely related to final state effects caused by an increase in the insulating properties of a thicker oxide layer.\cite{finster1985}

\begin{figure*}[]
    \centering
    \includegraphics[]{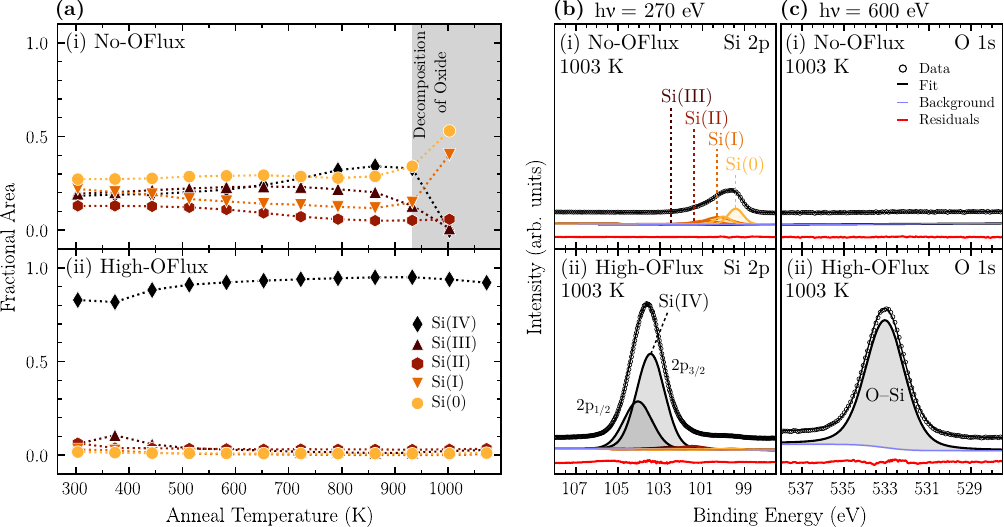}
    \subfiglabel{fgr:XPS_Anneal_Area}
    \subfiglabel{fgr:XPS_Anneal-Si2p}
    \subfiglabel{fgr:XPS_Anneal-O1s}
    \caption{
    SR-XPS data measured on samples No-OFlux and High-OFlux to monitor thermally-induced changes following UHV annealing.
    Samples were annealed in cycles; for each cycle, the temperature was increased by \qty{70}{\kelvin} up to \qty{1073}{\kelvin}.
    Spectra were recorded at room temperature following each annealing cycle.
    (a) shows the fractional area of each component in the Si~2p core-level spectrum, summing both 2p$_\text{3/2}$ and 2p$_\text{1/2}$ contributions, as a function of annealing temperature for samples (i) No-OFlux and (ii) High-OFlux.
    The oxide decomposes when annealing No-OFlux above \qty{933}{\kelvin}, as marked by the grey area.
    Thermal decomposition of the oxide is not observed for sample High-OFlux.
    These observations are demonstrated by SR-XPS spectra of the (b) Si~2p and (c) O~1s core-levels of (i) No-OFlux and (ii) High-OFlux following anneal to \qty{1003}{\kelvin}.
    Spectra were fit with a model and the fitting residuals are shown beneath each subplot.
    The number of data points plotted for each core-level has been reduced by a factor of 2 for clarity.
    The Y-axes have been scaled for all spectra to allow relative comparisons between samples by assuming an equal amount of C~1s intensity for clean HOPG.
    Residuals are plotted on the same scale.
    }
    \label{fgr:XPS_Anneal}
\end{figure*}

The evolution of the fractional area of all Si~2p core-level fit components as a function of deposition time for the samples described as No-OFlux, Low-OFlux, and High-OFlux are shown in Figure~\ref{fgr:O_vs_O2_Si2pArea}.
Depositing Si in \ce{O2} with no atomic O, to make the sample labelled No-OFlux, gives a final fraction of Si(IV) of \qty{<10}{\percent}.
Comparatively, for samples Low-OFlux and High-OFlux, where atomic O is co-deposited with Si, the final fraction of Si(IV) after \qty{3600}{\second} is \qty{60}{\percent} and \qty{80}{\percent}, respectively.

The results show a notable deviation from oxidation of Si on metallic substrates, as has been demonstrated for 2D silica synthesis.
2D silicas are commonly grown on metallic substrates, \textit{e.g.}, Ru(0001), via deposition of atomic Si in \ce{O2}, followed by post-annealing in \ce{O2}.\cite{loffler2010}
For Ru(0001) specifically, the success of this procedure is dependent on the formation of an oxygen pre-cover on the substrate, creating an $(2\times1)$O-Ru(0001) surface with which the incoming Si atoms can react.
Defect-free HOPG, however, does not exhibit dissociative adsorption of \ce{O2}.\cite{ni2012,nowakowski1992}
Moreover, \ce{O2} desorbs from the HOPG surface at temperatures above \qty{45}{\kelvin}.\cite{mohamed2022}
The lack of adsorbed \ce{O2} on HOPG could make oxidation of Si unfavourable, as has been demonstrated for sample No-OFlux, 
In contrast, exposing the HOPG surface to a beam of atomic O creates surface-situated epoxy groups, amongst other \ce{C-O} groups\cite{larciprete2012}, which could act as sites for oxidation of Si and nucleation of \ce{SiO_x} growth.
A similar approach has previously been reported for growth of metal oxide NPs on epitaxial graphene, allowing for homogeneous distribution of the formed NPs.\cite{johns2013}

Thus, co-deposition of Si and atomic O is an efficient method for growing a highly oxidised \ce{SiO_x} film on HOPG.
Moreover, adjusting the flux of atomic O allows for tuning the final oxidation of the resulting \ce{SiO_x}, allowing for controlled growths of highly oxidised or sub-oxidic films.
\ce{SiO} condensation in the ISM results in silicon oxide clusters with a variety of stoichiometries.\cite{reber2008}
The presented \ce{SiO_x} films with tunable stoichiometries thus provide an opportunity to account for the width of interstellar silicate stoichiometries.

\subsection{Stability of SiO$_\text{x}$ films against UHV annealing}\label{subsect:SiOxAnneal}
Thermally-induced changes in the chemical structure of samples No-OFlux and High-OFlux were monitored by measuring the Si~2p and O~1s core-levels at room temperature following anneal in UHV.
The samples were annealed in steps of \qty{70}{\kelvin} from \qtyrange[range-units=single]{373}{1073}{\kelvin}.
The temperature was kept stable for \qty{1200}{\second} at each temperature step.
The results following anneal of a sample produced without atomic O, namely sample No-OFlux and sample High-OFlux, produced using atomic O, are presented in Figure~\ref{fgr:XPS_Anneal}.

Annealing of sample No-OFlux results in complete decomposition of the oxide above \qty{933}{\kelvin}, Figure~\ref{fgr:XPS_Anneal_Area}(i).
Prior to decomposition, annealing leads to segregation of the Si into primarily Si(0) and Si(IV) species, as shown in Figure~\ref{fgr:XPS_Anneal_Area}(i), with Si(IV) most abundant at \qty{863}{\kelvin}.
The Si~2p and O~1s core-level spectra, shown in Figures~\ref{fgr:XPS_Anneal-Si2p}(i) and \ref{fgr:XPS_Anneal-O1s}(i), respectively, confirms the complete desorption of the oxide after annealing the sample No-OFlux to \qty{1003}{\kelvin}.
This behaviour is comparable to the thermal decomposition of the oxide layer on bulk Si surfaces through the generation of volatile \ce{SiO} via the reaction \ce{SiO2(s) + Si(s) -> 2SiO(g)}.\cite{lander1962,streit1987} 
Moreover, the majority of the elemental silicon from sample No-OFlux was observed to desorb in the decomposition reaction.

Figure~\ref{fgr:XPS_Anneal_Area}(ii) plots the fractional areas of the individual Si~2p components as a function of annealing temperature for sample High-OFlux.
The lower Si oxidation states drop in fractional area during anneal, while the Si(IV) component constitutes a increasing fraction of the \ce{SiO_x} film.
The Si(IV) component reaches a maximum fractional area of \qty{95}{\percent} at \qty{933}{\kelvin} before dropping slightly to \qty{92}{\percent} at \qty{1073}{\kelvin}.
Films could not be heated above \qty{1073}{\kelvin} in the experimental chamber.
This suggests that thermal decomposition of the film begins above \qty{933}{\kelvin}.
At \qty{933}{\kelvin}, sample High-OFlux is stoichiometrically very close to \ce{SiO2}.

Figure~\ref{fgr:XPS_Anneal-Si2p}(ii) shows the Si~2p core-level spectrum captured following an anneal to \qty{1003}{\kelvin}.
The Si~2p core-level signal exhibits a dominant Si(IV) component at \qty{103.5}{\electronvolt} for the Si~2p$_\text{3/2}$ peak.
The lower Si oxidation states, \textit{i.e.}, Si(0 $\leq$ n $\leq$ III), are noticeably suppressed after anneal to \qty{1003}{\kelvin} compared to the as-grown film, presented in Figure~\ref{fgr:O_vs_O2}.

The O~1s core-level spectrum for sample High-OFlux after anneal to \qty{1003}{\kelvin} is shown in Figure~\ref{fgr:XPS_Anneal-O1s}(ii).
The signal can, similar to the data in Figure~\ref{fgr:O_vs_O2_O1s}, be fitted to a single component at \qty{533.1}{\electronvolt}, corresponding to \ce{O-Si} bonds.
The FWHM of the peak decreases from \qty{2.30}{\electronvolt} for the as-grown film to \qty{2.20}{\electronvolt} following anneal to \qty{653}{\kelvin} and above.
A decreasing FWHM is tied to a decrease in the diversity of the chemical states of O.
As such, annealing the \ce{SiO_x} film leads to an increasingly uniform chemical structure.

\begin{figure*}[ht]
    \centering
    \includegraphics[]{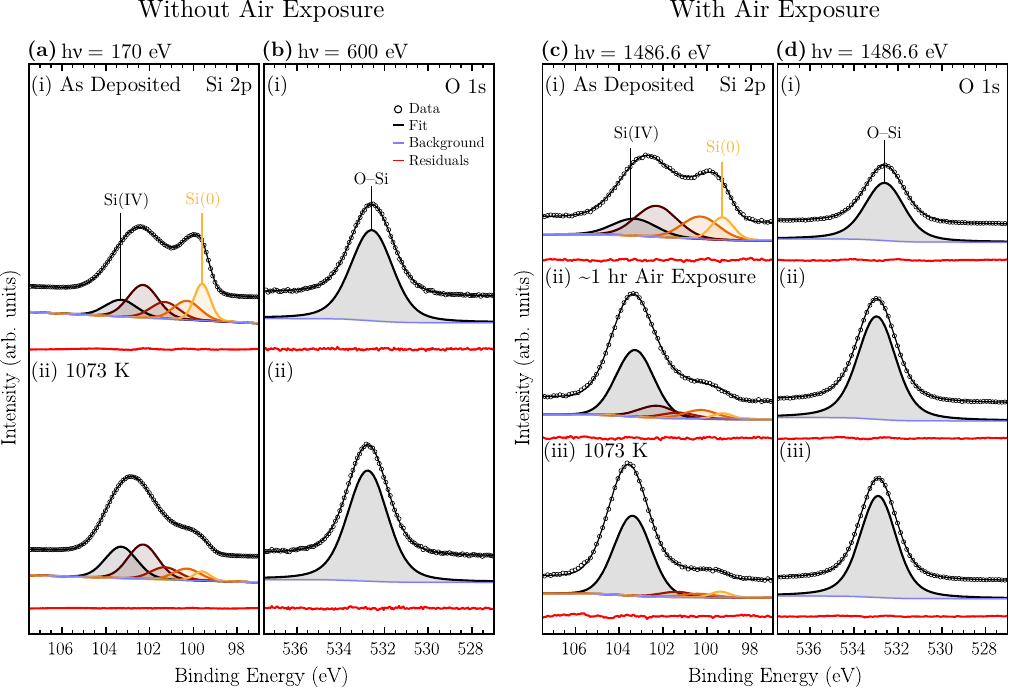}
    \subfiglabel{fgr:XPS_NoAirExposure_Si2p}
    \subfiglabel{fgr:XPS_NoAirExposure_O1s}
    \subfiglabel{fgr:XPS_AirExposure_Si2p}
    \subfiglabel{fgr:XPS_AirExposure_O1s}
    \caption{XPS data recorded from suboxidised \ce{SiO_x} samples (a,b) without and (c,d) with exposure to air prior to annealing at \qty{1073}{\kelvin}.
    (a) Si~2p and (b) O~1s core-levels of \ce{SiO_x} on HOPG annealed without prior air exposure are shown on the left.
    Spectra were recorded for the films (i) as deposited and (ii) after annealing to \qty{1073}{\kelvin} in UHV.
    Annealing to \qty{1073}{\kelvin} leads to a loss in the Si(0) component, while the Si(IV) component is enhanced.
    A small reduction in the overall Si signal is observed.
    Similarly, (c) Si~2p and (d) O~1s core-levels of \ce{SiO_x} on HOPG with air exposure prior to annealing are shown on the right.
    Spectra for the \ce{SiO_x} film (i) as deposited, (ii) after \qty{1}{\hour} of air exposure, and (iii) after annealing to \qty{1073}{\kelvin} are shown for both core-levels.
    Air exposure and subsequent annealing leads to oxidation of the Si, resulting in a dominant Si(IV) component.
    No reduction in the overall Si signal is observed.
    In the Si~2p core-level spectra, the Si(IV) and Si(0) components have been highlighted at the respective BEs of \qty{103.5}{\electronvolt} and \qty{99.5}{\electronvolt}.
    Data are shown as black circles and have been fit with a model; the fitting residuals are shown beneath each subplot.
    Backgrounds for each spectrum are plotted as blue lines.
    For clarity, only Si~2p$_\text{3/2}$ components are plotted in the Si~2p core-level spectrum.
    Similarly, the number of plotted data points has been reduced by a factor of 5 for (a) and a factor of 2 for (b)-(d).
    The intensities of the Y-axes have been adjusted for qualitative comparison; differences in kinetic energy, photoionisation cross sections, and photon energies disallow quantitative comparisons between subfigures.}
    \label{fgr:XPS_AirExposure}
\end{figure*}

We conclude that the growth of \ce{SiO_x} samples using atomic O produces films with better resilience against thermal decomposition.
We relate this stability to either i) the specific morphology of the \ce{SiO_x} film achieved when atomic O is incorporated or to ii) the higher level of Si-oxidation obtained when co-depositing with atomic O, wherein the lack of Si(0) prohibits generation of SiO.
To test these two hypotheses, two films of \ce{SiO_x} were prepared using much lower fluence of atomic O, giving intentional suboxidation of Si and a thickness of approximately \qty{4}{\angstrom}, and then these samples were annealed either with or without prior exposure to air.
Air exposure represents a saturation of the \ce{SiO_x} film in \ce{O2}.
XPS measurements were used to characterise the degree of oxidation induced by air exposure and the corresponding effect that this had on the stability of the film in response to annealing.
The results are presented in Figure~\ref{fgr:XPS_AirExposure}.

Figure~\ref{fgr:XPS_NoAirExposure_Si2p} \& \ref{fgr:XPS_NoAirExposure_O1s} show XPS measurements of the photoelectron flux at the Si~2p and O~1s core-levels of a sub-oxidic \ce{SiO_x} film annealed to \qty{1073}{\kelvin} without prior air exposure.
The film of \ce{SiO_x} is initially sub-oxidic \ce{SiO_x}, with a distribution of Si oxidation states (Figure~\ref{fgr:XPS_NoAirExposure_Si2p}~(i)) and a small O~1s peak centered at a binding energy of \qty{532.6}{\electronvolt} (Figure~\ref{fgr:XPS_NoAirExposure_O1s}~(i)).
The O~1s core-level signal is fitted to a single \ce{O-Si} component.
Annealing of the film to \qty{1073}{\kelvin} in UHV results in an enhancement of the Si(IV) component in the Si~2p core-level (Figure~\ref{fgr:XPS_NoAirExposure_Si2p}~(ii)) and simultaneously, the Si(0) is heavily attenuated.
There is an overall \qty{10}{\percent} loss of Si~2p intensity.
We note that the sub-oxidic \ce{SiO_x} film exhibits resilience against thermal decomposition and, rather than desorbing, moves toward a higher overall Si oxidation state.
This is further supported by the O~1s core-level spectrum (Figure~\ref{fgr:XPS_NoAirExposure_O1s}~(ii)), wherein the \ce{O-Si} component is observed to shift to \qty{532.8}{\electronvolt}, indicating a thicker oxide layer.\cite{finster1985} 
The suboxide film exhibits resilience toward thermal decomposition, in stark contrast to sample No-OFlux where the oxide decomposed completely at \qty{1003}{\kelvin}, as discussed in Figure~\ref{fgr:XPS_Anneal}.

Figure~\ref{fgr:XPS_AirExposure_Si2p} \& \ref{fgr:XPS_AirExposure_O1s} show XPS measurements of the photoelectron spectra at the Si~2p and O~1s core-levels for a sub-oxidic \ce{SiO_x} film, which was exposed to air prior to annealing to \qty{1073}{\kelvin} in UHV.
The \ce{SiO_x} sample, Figure~\ref{fgr:XPS_AirExposure_Si2p}~(i), has a composition of Si-states comparable to that of the sample presented in Figure~\ref{fgr:XPS_NoAirExposure_Si2p} \& \ref{fgr:XPS_NoAirExposure_O1s}, with the O~1s peak in Figure~\ref{fgr:XPS_AirExposure_O1s}~(i) centered at a binding energy of \qty{532.6}{\electronvolt}.
Air exposure for approximately one hour results in oxidation of the film and leads to an increase in intensity of the Si(IV) state, Figure~\ref{fgr:XPS_AirExposure_Si2p}~(ii).
In addition, the O~1s peak signal visibly increases and shifts to \qty{533}{\electronvolt}, as expected during silicon oxidation, Figure~\ref{fgr:XPS_AirExposure_O1s}~(ii).
Further enhancement of the Si(IV) state signal is observed after annealing to \qty{1073}{\kelvin}, while the lower oxidation states are lost, Figure~\ref{fgr:XPS_AirExposure_Si2p}~(iii).
The O~1s peak shifts slightly to \qty{532.9}{\electronvolt} (Figure~\ref{fgr:XPS_AirExposure_O1s}~(iii)).
Taken together, Figure~\ref{fgr:XPS_AirExposure_Si2p} \& \ref{fgr:XPS_AirExposure_O1s} demonstrate that air exposure oxidises the surface of the sub-oxidic \ce{SiO_x} film and annealing results in further oxidation of the Si, likely by bond rearrangements.
Even without complete oxidation via air exposure, \ce{SiO_x} films grown on HOPG by co-depositing Si with atomic O obtain a morphology which prevents thermal decomposition of the oxide at \qty{1073}{\kelvin}.

\begin{figure}[]
    \centering
    \includegraphics[]{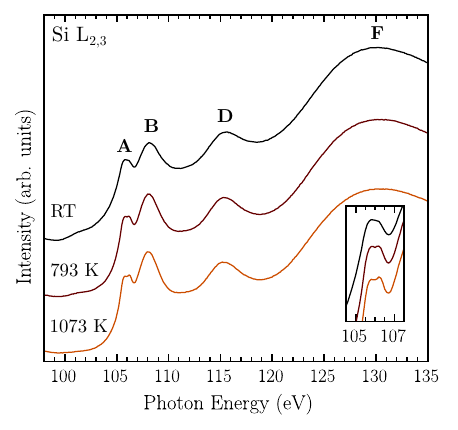}
    \subfiglabel{fgr:NEXAFS_Anneal-SiLEdge}
    \caption{NEXAFS spectra of the Si L$_{2,3}$-edge for a \ce{SiO_x}/HOPG sample prepared using atomic O.
    Data were recorded after deposition (top), and after annealing to \qty{793}{\kelvin} (middle) and \qty{1073}{\kelvin} (bottom).
    The peak naming convention follows that of \citeauthor{chaboy1995theoretical}, see comments in main texts.
    Inset; Structure in peak A related to spin-orbit splitting of Si~2p is revealed in peak A upon annealing. Data are offset on the y-axis for clarity.}
    \label{fgr:NEXAFS_Anneal}
\end{figure}

NEXAFS spectra of the Si L$_{2,3}$-edge for a \ce{SiO_x} film grown by co-depositing atomic Si and O are shown in Figure~\ref{fgr:NEXAFS_Anneal}. 
Spectra were recorded immediately after growth, labelled "RT", and following anneal to \qty{793}{\kelvin} and then \qty{1073}{\kelvin}. 
Data were recorded at room temperature.
Several peaks constitute the absorption edge structure and indicate the formation of \ce{SiO2}. 
Following the naming convention used for \ce{SiO2} by \citeauthor{chaboy1995theoretical}\cite{chaboy1995theoretical}, peaks are observed at \qty{106}{\electronvolt}~(A), \qty{108}{\electronvolt}~(B), \qty{115}{\electronvolt}~(D), and \qty{130}{\electronvolt}~(F).
The reader is referred to \citeauthor{li1993high}\cite{li1993high} and \citeauthor{chaboy1995theoretical}\cite{chaboy1995theoretical} for a full discussion on the physical origin of these peaks.
The rise in intensity in the data measured immediately after growth, from \qty{100}{\electronvolt} to \qty{105}{\electronvolt}, likely originates from the presence of non-oxidised Si, \textit{i.e.}, Si(0), which has Si L$_{2,3}$-edge peaks in this energy range.\cite{Brown1977}
The absence of this rise in the data captured after annealing to \qty{1073}{\kelvin} signifies that this Si has either been oxidised or desorbed, in agreement with the XPS data presented in Figures~\ref{fgr:XPS_Anneal} and \ref{fgr:XPS_NoAirExposure_Si2p}.
Annealing also leads to the appearance of structure in peak A, with more defined peaks as the annealing temperature increases. 
This structure originates from spin-orbit splitting of Si~2p.\cite{li1993high}
Such structure in the absorption edge features has previously been reported for samples of crystalline \ce{SiO2}, \textit{e.g.}, $\alpha$-quartz, while glassy \ce{SiO2} does not exhibit the splitting.\cite{garvie1999}
In summary, the XPS and NEXAFS data indicate that \ce{SiO_x}/HOPG produced using atomic O during co-deposition evolves toward a more well-defined chemical structure with some long range order, following anneal under UHV conditions.

\subsection{Complete coverage of the HOPG surface by SiO$_x$ NPs}
\begin{figure*}[]
    \centering
    \includegraphics[]{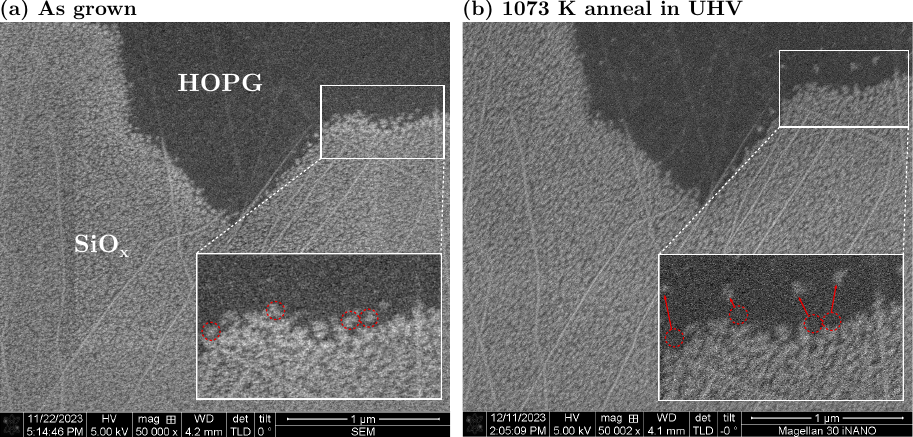}
    \subfiglabel{fgr:SEM-RT_SEM}
    \subfiglabel{fgr:SEM-1073K_SEM}
    \caption{SEM images of \ce{SiO_x} on HOPG showing a boundary between a region of bare HOPG and \ce{SiO_x}.
    Refer to the main text for an explanation on the physical origin of this region.
    In (a), the as-grown sample is shown, while in (b), the sample is shown after annealing to \qty{1073}{\kelvin} in UHV. 
    The \ce{SiO_x} film has a macroscopic homogeneously distributed, nanoparticulate morphology.
    The morphology does not seem to change during anneal, but some migration of \ce{SiO_x} clusters into the bare HOPG region is observed, as demonstrated in the insets to (a) and (b).}
    \label{fgr:SEM}
\end{figure*}

The morphology of the \ce{SiO_x} films on HOPG was investigated using SEM. 
A \ce{SiO_x}/HOPG sample, grown to a thickness of \qty{4}{\angstrom}, was taken out of UHV and transferred to a different experimental chamber for SEM imaging, resulting in approximately \qty{30}{\min} air exposure. 
After this initial imaging experiment, the sample was reintroduced to UHV, degassed in \qty{100}{\kelvin} steps and finally annealed to \qty{1073}{\kelvin}.
The sample was imaged again, after annealing in UHV, to assess changes to the \ce{SiO_x} film morphology.
SEM images of the same region of \ce{SiO_x} film as-grown and following annealing at \qty{1073}{\kelvin} are presented in Figure~\ref{fgr:SEM}.
An image of the as-grown \ce{SiO_x} film is shown in Figure~\ref{fgr:SEM-RT_SEM}.
The image is captured at the site of a lifted flake of graphite to highlight the contrast difference between the bare HOPG and the \ce{SiO_x} film.
Large scale SEM images of the lifted flake can be seen in Figure~S2.

The darker area towards the top of Figure~\ref{fgr:SEM-RT_SEM} is attributed to bare HOPG, based on imaging of a clean HOPG reference sample.
The area populated by white islands is then assigned to the \ce{SiO_x} film.
Areas of comparable contrast were not found on the clean HOPG reference sample.
Images captured on large, flat areas away from the lifted flake of the HOPG substrate exhibit a continuous distribution of the \ce{SiO_x} film with same morphology, demonstrating growth of \ce{SiO_x} on a macroscopic homogeneous level across the exposed substrate surface (Figure~S2).
The lack of \ce{SiO_x} growth underneath the graphite flake suggests that the \ce{SiO_x} film grows only on regions in line-of-sight of the Si and O beams.
This concurrently demonstrates that the co-deposited Si and O do not penetrate the upper HOPG layers, resulting only in surface growth of \ce{SiO_x}.
Straight lines are also observed throughout the \ce{SiO_x}-covered areas.
We propose that these lines are grain-boundaries of the HOPG and have increased affinity for \ce{SiO_x} growth, explaining their enhanced contrast and continuity.
Overall, the as-grown \ce{SiO_x} film appears as an interconnected high surface area NP network, homogeneously distributed on the HOPG surface within line-of-sight of the atomic beams.

Figure~\ref{fgr:SEM-1073K_SEM} shows an SEM image taken after annealing the film to \qty{1073}{\kelvin} in UHV.
The region imaged is the same as in Figure~\ref{fgr:SEM-RT_SEM}.
Little change in morphology is observed; indeed, as shown in the insets of Figure~\ref{fgr:SEM-RT_SEM} \& \ref{fgr:SEM-1073K_SEM}, some \ce{SiO_x} particles retain their position post-annealing. 
Thus the chemical changes induced by annealing do not lead to observable changes of morphology at the nanoscale, visible using SEM. 
This suggests that the chemical changes observed with XPS after annealing happen internally in individual \ce{SiO_x} clusters. 
However, isolated \ce{SiO_x} clusters are observed in the bare HOPG region after the annealing step, suggesting thermally induced cluster migration.
Clusters that appear to have migrated during the annealing are highlighted with red dotted circles in the inset of Figure~\ref{fgr:SEM-RT_SEM}.
The suggested migration path of the clusters are shown with arrows in the inset to Figure~\ref{fgr:SEM-1073K_SEM}.
It is likely that this migration is only possible because the clusters were originally separate from the interconnected matrix of \ce{SiO_x} and situated next to a region of bare HOPG.

\begin{figure*}[h]
    \centering
    \includegraphics[width=17.1cm]{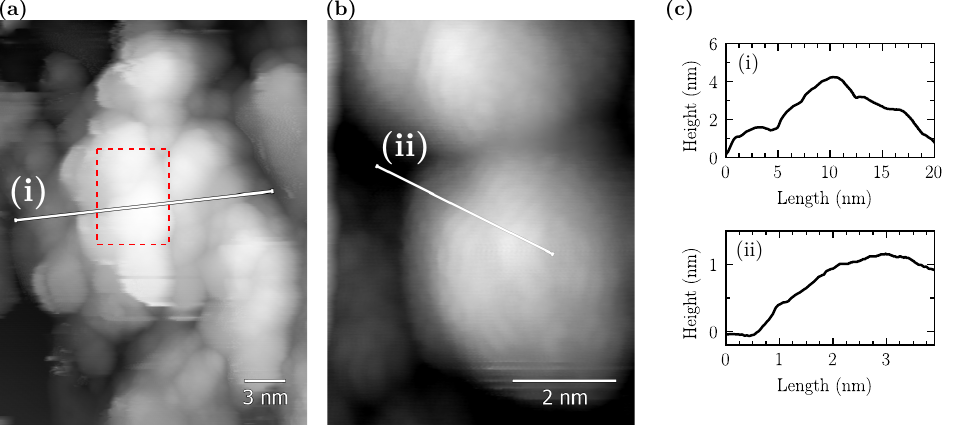}
    \subfiglabel{fgr:STM-largescale}
    \subfiglabel{fgr:STM-smallscale}
    \subfiglabel{fgr:STM-XPS_data}
    \caption{STM images of \ce{SiO_x} nanoclusters grown on HOPG via codeposition of atomic Si and O, and then annealed to \qty{1043}{\kelvin}. In (a) (V$_\text{t}=\qty{2996.6}{\milli\volt}$, I$_\text{t}=\qty{0.270}{\nano\ampere}$), a line scan across the clusters gives an apparent height of \qty{4}{\nano\meter}, shown in (c - i). (b) (V$_\text{t}=\qty{502.3}{\milli\volt}$, I$_\text{t}=\qty{0.060}{\nano\ampere}$) shows a high-resolution image of the inset in (a), with a line scan showing that the cluster has a radius of \qty{\sim2.5}{\nano\meter}, shown in (c - ii).}
    \label{fgr:STM}
\end{figure*}

To further investigate the structure of the \ce{SiO_x} clusters observed with SEM, Si and O were codeposited on clean HOPG, annealed to \qty{1043}{\kelvin} and the resulting film was imaged with STM. 
Representative STM images are shown in Figure~\ref{fgr:STM}.
STM characterisation reveals that the diffuse nanoparticulate film imaged with SEM and visible in Figure~\ref{fgr:SEM} is itself composed of smaller sized clusters.
A line scan across the \ce{SiO_x} growth represented in Figure~\ref{fgr:STM-largescale} demonstrates a particle width of approximately \qty{20}{\nano\meter} with an apparent height of approximately \qty{4}{\nano\meter}. 
The width of this growth is equivalent to some of the smallest particles visible in Figure~\ref{fgr:SEM}. 
This particle is surrounded by low valleys of exposed or partially covered HOPG areas, which is in good agreement with the structures observed with SEM presented in Figure~\ref{fgr:SEM}.
A high-resolution image of the area outlined in Figure~\ref{fgr:STM-largescale} is presented in Figure~\ref{fgr:STM-smallscale}.
This image reveals a hemi-spherical \ce{SiO_x} cluster with a radius of approximately \qty{2.5}{\nano\meter}.

\section{Conclusion}
We have demonstrated the growth of a thin film of nanoparticulate clusters of \ce{SiO_x} with a macroscopic homogeneous distribution on a HOPG substrate. 
The \ce{SiO_x} film was sufficiently conducting to be used for classic surface science characterisation using photo-electron spectroscopy, X-ray absorption spectrosocpy and scanning tunnelling microscopy. 
These analyses, combined with \textit{ex situ} analysis via SEM, show that \ce{SiO_x} films grown by co-deposition of atomic Si and atomic O, allows for tuning of the average oxidation state of Si throughout the film, produces a high surface area, macroscopic homogeneous film of interconnected NPs on the HOPG substrate surface, and results in \ce{SiO_x} films that exhibit high resilience to thermal decomposition, tested up to \qty{1073}{\kelvin}.
The \ce{SiO_x} films are composed of nanoscale \ce{SiO_x} clusters with dimensions of \qty{\sim2.5}{\nano\meter}, that can be grown \textit{in situ} to predominantly contain Si in the Si(IV) state, or, in case of O-poor growths, a mixture of Si sub-oxides. 
Annealing is reported to lead to an increase in the oxidation state of the film, via loss of the lower oxidised states of Si, but this was not observed to produce morphology changes to the film when characterised using SEM.

We have presented a system for growing sub-oxidic or fully oxidised \ce{SiO_x} films on an chemically inert substrate.
The proposed co-deposition method can be adapted to include co-deposition with other elements and, in the context of laboratory astrochemistry, inclusion of Fe or Mg atoms will be of particular importance.
Specifically, it is of interest to study how such metal inclusions affect the silicate structure, as well as its spectroscopic signature in the infrared, XPS, and more.
Moreover, the substrates have potential uses as model systems for high surface area silica support materials relevant to the study of industrial catalyst materials.
The thin, nanoscale and conducting substrates demonstrated here could be used to conduct surface science studies on the catalytic activity of \ce{SiO_x} materials as a function of metal-atom inclusion. 
Studies are currently underway to investigate the possibilities and limitations of the method in regards to these areas of interest, particularly as they relate to astrochemistry.

\section*{Author contributions}
\textbf{Steffen Friis Holleufer:} Conceptualization, Validation, Formal Analysis, Investigation, Methodology, Data Curation, Visualization, Writing - Original Draft.
\textbf{Alfred Hopkinson:} Investigation.
\textbf{Duncan S. Sutherland:} Investigation, Resources.
\textbf{Zheshen Li:} Investigation, Resources.
\textbf{Jeppe Vang Lauritsen:} Supervision.
\textbf{Liv Hornekær:} Conceptualization, Resources, Writing - Review and Editing, Supervision, Project Administration, Funding Acquisition.
\textbf{Andrew Cassidy:} Conceptualization, Writing - Review and Editing, Supervision, Project Administration.

\section*{Conflicts of interest}
There are no conflicts to declare.

\section*{Data availability}
All data will be made available and hosted on the Zenodo database. A DOI to this archive will be made available after the paper is accepted  for publication. 

\section*{Acknowledgements}
The work is supported by the Danish National Research Foundation through the Center of Excellence "InterCat" (grant agreement no. DNRF150).
We acknowledge beam time received at ASTRID2 on the AU-Matline beamline under project numbers ISA-23-1010 and ISA-24-1221.



\balance


\bibliography{Paper}

\providecommand*{\mcitethebibliography}{\thebibliography}
\csname @ifundefined\endcsname{endmcitethebibliography}
{\let\endmcitethebibliography\endthebibliography}{}
\begin{mcitethebibliography}{67}
\providecommand*{\natexlab}[1]{#1}
\providecommand*{\mciteSetBstSublistMode}[1]{}
\providecommand*{\mciteSetBstMaxWidthForm}[2]{}
\providecommand*{\mciteBstWouldAddEndPuncttrue}
  {\def\EndOfBibitem{\unskip.}}
\providecommand*{\mciteBstWouldAddEndPunctfalse}
  {\let\EndOfBibitem\relax}
\providecommand*{\mciteSetBstMidEndSepPunct}[3]{}
\providecommand*{\mciteSetBstSublistLabelBeginEnd}[3]{}
\providecommand*{\EndOfBibitem}{}
\mciteSetBstSublistMode{f}
\mciteSetBstMaxWidthForm{subitem}
{(\emph{\alph{mcitesubitemcount}})}
\mciteSetBstSublistLabelBeginEnd{\mcitemaxwidthsubitemform\space}
{\relax}{\relax}

\bibitem[Van~Steenberg and Shull(1988)]{vansteenberg1988}
M.~E. Van~Steenberg and J.~M. Shull, \emph{ApJ}, 1988, \textbf{330}, 942--963\relax
\mciteBstWouldAddEndPuncttrue
\mciteSetBstMidEndSepPunct{\mcitedefaultmidpunct}
{\mcitedefaultendpunct}{\mcitedefaultseppunct}\relax
\EndOfBibitem
\bibitem[Tielens(1998)]{tielens1998}
A.~G. G.~M. Tielens, \emph{ApJ}, 1998, \textbf{499}, 267--272\relax
\mciteBstWouldAddEndPuncttrue
\mciteSetBstMidEndSepPunct{\mcitedefaultmidpunct}
{\mcitedefaultendpunct}{\mcitedefaultseppunct}\relax
\EndOfBibitem
\bibitem[Draine(2003)]{draine2003}
B.~T. Draine, \emph{ARA\&A}, 2003, \textbf{41}, 241--289\relax
\mciteBstWouldAddEndPuncttrue
\mciteSetBstMidEndSepPunct{\mcitedefaultmidpunct}
{\mcitedefaultendpunct}{\mcitedefaultseppunct}\relax
\EndOfBibitem
\bibitem[Henning(2010)]{henning2010}
T.~Henning, \emph{ARA\&A}, 2010, \textbf{48}, 21--46\relax
\mciteBstWouldAddEndPuncttrue
\mciteSetBstMidEndSepPunct{\mcitedefaultmidpunct}
{\mcitedefaultendpunct}{\mcitedefaultseppunct}\relax
\EndOfBibitem
\bibitem[Zhukovska \emph{et~al.}(2016)Zhukovska, Dobbs, Jenkins, and Klessen]{zhukovska2016}
S.~Zhukovska, C.~Dobbs, E.~B. Jenkins and R.~S. Klessen, \emph{ApJ}, 2016, \textbf{831}, 147 (15pp)\relax
\mciteBstWouldAddEndPuncttrue
\mciteSetBstMidEndSepPunct{\mcitedefaultmidpunct}
{\mcitedefaultendpunct}{\mcitedefaultseppunct}\relax
\EndOfBibitem
\bibitem[Zhukovska \emph{et~al.}(2018)Zhukovska, Henning, and Dobbs]{zhukovska2018}
S.~Zhukovska, T.~Henning and C.~Dobbs, \emph{ApJ}, 2018, \textbf{857}, 94 (12pp)\relax
\mciteBstWouldAddEndPuncttrue
\mciteSetBstMidEndSepPunct{\mcitedefaultmidpunct}
{\mcitedefaultendpunct}{\mcitedefaultseppunct}\relax
\EndOfBibitem
\bibitem[Demyk(2011)]{demyk2011}
K.~Demyk, \emph{EPJ Web Conf.}, 2011, \textbf{18}, 03001\relax
\mciteBstWouldAddEndPuncttrue
\mciteSetBstMidEndSepPunct{\mcitedefaultmidpunct}
{\mcitedefaultendpunct}{\mcitedefaultseppunct}\relax
\EndOfBibitem
\bibitem[Hollenbach and Salpeter(1971)]{hollenbach1971}
D.~Hollenbach and E.~E. Salpeter, \emph{ApJ}, 1971, \textbf{163}, 155--164\relax
\mciteBstWouldAddEndPuncttrue
\mciteSetBstMidEndSepPunct{\mcitedefaultmidpunct}
{\mcitedefaultendpunct}{\mcitedefaultseppunct}\relax
\EndOfBibitem
\bibitem[Vidali \emph{et~al.}(2009)Vidali, Li, Roser, and Badman]{vidali2009}
G.~Vidali, L.~Li, J.~E. Roser and R.~Badman, \emph{ASR}, 2009, \textbf{43}, 1291--1298\relax
\mciteBstWouldAddEndPuncttrue
\mciteSetBstMidEndSepPunct{\mcitedefaultmidpunct}
{\mcitedefaultendpunct}{\mcitedefaultseppunct}\relax
\EndOfBibitem
\bibitem[Vidali(2013)]{vidali2013}
G.~Vidali, \emph{Chem. Rev.}, 2013, \textbf{113}, 8762--8782\relax
\mciteBstWouldAddEndPuncttrue
\mciteSetBstMidEndSepPunct{\mcitedefaultmidpunct}
{\mcitedefaultendpunct}{\mcitedefaultseppunct}\relax
\EndOfBibitem
\bibitem[Suhasaria and Mennella(2021)]{suhasaria2021}
T.~Suhasaria and V.~Mennella, \emph{Front. Astron. Space Sci.}, 2021, \textbf{8}, 655883\relax
\mciteBstWouldAddEndPuncttrue
\mciteSetBstMidEndSepPunct{\mcitedefaultmidpunct}
{\mcitedefaultendpunct}{\mcitedefaultseppunct}\relax
\EndOfBibitem
\bibitem[Serra-Peralta \emph{et~al.}(2022)Serra-Peralta, Dominguez-Dalmases, and Rimola]{serraperalta2022}
M.~Serra-Peralta, C.~Dominguez-Dalmases and A.~Rimola, \emph{Phys. Chem. Chem. Phys.}, 2022, \textbf{24}, 28381--28393\relax
\mciteBstWouldAddEndPuncttrue
\mciteSetBstMidEndSepPunct{\mcitedefaultmidpunct}
{\mcitedefaultendpunct}{\mcitedefaultseppunct}\relax
\EndOfBibitem
\bibitem[McClure \emph{et~al.}(2023)McClure, Rocha, Pontoppidan, Crouzet, Chu, Dartois, Lamberts, Noble, Pendleton, Perotti, Qasim, Rachid, Smith, Sun, Beck, Boogert, Brown, Caselli, Charnley, Cuppen, Dickinson, Drozdovskaya, Egami, Erkal, Fraser, Garrod, Harsono, Ioppolo, Jim\'{e}nez-Serra, Jin, Jørgensen, Kristensen, Lis, McCoustra, McGuire, Melnick, \"{O}berg, Palumbo, Shimonishi, Sturm, van Dishoeck, and Linnartz]{mcclure2024}
M.~K. McClure, W.~R.~M. Rocha, K.~M. Pontoppidan, N.~Crouzet, L.~E.~U. Chu, E.~Dartois, T.~Lamberts, J.~A. Noble, Y.~J. Pendleton, G.~Perotti, D.~Qasim, M.~G. Rachid, Z.~L. Smith, F.~Sun, T.~L. Beck, A.~C.~A. Boogert, W.~A. Brown, P.~Caselli, S.~B. Charnley, H.~M. Cuppen, H.~Dickinson, M.~N. Drozdovskaya, E.~Egami, J.~Erkal, H.~Fraser, R.~T. Garrod, D.~Harsono, S.~Ioppolo, I.~Jim\'{e}nez-Serra, M.~Jin, J.~K. Jørgensen, L.~E. Kristensen, D.~C. Lis, M.~R.~S. McCoustra, B.~A. McGuire, G.~J. Melnick, K.~I. \"{O}berg, M.~E. Palumbo, T.~Shimonishi, J.~A. Sturm, E.~F. van Dishoeck and H.~Linnartz, \emph{Nat. Astron.}, 2023, \textbf{7}, 431--443\relax
\mciteBstWouldAddEndPuncttrue
\mciteSetBstMidEndSepPunct{\mcitedefaultmidpunct}
{\mcitedefaultendpunct}{\mcitedefaultseppunct}\relax
\EndOfBibitem
\bibitem[Voshchinnikov and Henning(2010)]{voshchinnikov2010}
N.~V. Voshchinnikov and T.~Henning, \emph{A\&A}, 2010, \textbf{517}, A45\relax
\mciteBstWouldAddEndPuncttrue
\mciteSetBstMidEndSepPunct{\mcitedefaultmidpunct}
{\mcitedefaultendpunct}{\mcitedefaultseppunct}\relax
\EndOfBibitem
\bibitem[He \emph{et~al.}(2022)He, Luo, Doddipatla, Yang, Millar, Sun, and Kaiser]{he2022}
C.~He, Y.~Luo, S.~Doddipatla, Z.~Yang, T.~J. Millar, R.~Sun and R.~I. Kaiser, \emph{Phys. Chem. Chem. Phys.}, 2022, \textbf{24}, 19761\relax
\mciteBstWouldAddEndPuncttrue
\mciteSetBstMidEndSepPunct{\mcitedefaultmidpunct}
{\mcitedefaultendpunct}{\mcitedefaultseppunct}\relax
\EndOfBibitem
\bibitem[Reber \emph{et~al.}(2008)Reber, Paranthaman, Clayborne, Khanna, and Castleman]{reber2008}
A.~C. Reber, S.~Paranthaman, P.~A. Clayborne, S.~N. Khanna and A.~W.~J. Castleman, \emph{ACS Nano}, 2008, \textbf{2}, 1729--1737\relax
\mciteBstWouldAddEndPuncttrue
\mciteSetBstMidEndSepPunct{\mcitedefaultmidpunct}
{\mcitedefaultendpunct}{\mcitedefaultseppunct}\relax
\EndOfBibitem
\bibitem[Henning \emph{et~al.}(2017)Henning, J\"{a}ger, Rouillé, Fulvio, and Krasnokutski]{henning2017}
T.~Henning, C.~J\"{a}ger, G.~Rouillé, D.~Fulvio and S.~A. Krasnokutski, \emph{Proc. IAU Symp.}, 2017, \textbf{13}, 312--319\relax
\mciteBstWouldAddEndPuncttrue
\mciteSetBstMidEndSepPunct{\mcitedefaultmidpunct}
{\mcitedefaultendpunct}{\mcitedefaultseppunct}\relax
\EndOfBibitem
\bibitem[Li and Draine(2002)]{li2002}
A.~Li and B.~T. Draine, \emph{ApJ}, 2002, \textbf{564}, 803--812\relax
\mciteBstWouldAddEndPuncttrue
\mciteSetBstMidEndSepPunct{\mcitedefaultmidpunct}
{\mcitedefaultendpunct}{\mcitedefaultseppunct}\relax
\EndOfBibitem
\bibitem[Sargent \emph{et~al.}(2009)Sargent, Forrest, Tayrien, McClure, Li, Basu, Manoj, Watson, Bohac, and Furlan]{sargent2009}
B.~A. Sargent, W.~J. Forrest, C.~Tayrien, M.~K. McClure, A.~Li, A.~R. Basu, P.~Manoj, D.~M. Watson, C.~J. Bohac and E.~Furlan, \emph{ApJ}, 2009, \textbf{690}, 1193--1207\relax
\mciteBstWouldAddEndPuncttrue
\mciteSetBstMidEndSepPunct{\mcitedefaultmidpunct}
{\mcitedefaultendpunct}{\mcitedefaultseppunct}\relax
\EndOfBibitem
\bibitem[Draine and Hensley(2021)]{Draine2021}
B.~T. Draine and B.~S. Hensley, \emph{ApJ}, 2021, \textbf{909}, 94\relax
\mciteBstWouldAddEndPuncttrue
\mciteSetBstMidEndSepPunct{\mcitedefaultmidpunct}
{\mcitedefaultendpunct}{\mcitedefaultseppunct}\relax
\EndOfBibitem
\bibitem[Hensley and Draine(2023)]{Hensley2023}
B.~S. Hensley and B.~T. Draine, \emph{ApJ}, 2023, \textbf{948}, 55\relax
\mciteBstWouldAddEndPuncttrue
\mciteSetBstMidEndSepPunct{\mcitedefaultmidpunct}
{\mcitedefaultendpunct}{\mcitedefaultseppunct}\relax
\EndOfBibitem
\bibitem[Schneider and Beall~Fowler(1976)]{schneider1976}
P.~M. Schneider and W.~Beall~Fowler, \emph{Phys. Rev. Lett.}, 1976, \textbf{36}, 425--428\relax
\mciteBstWouldAddEndPuncttrue
\mciteSetBstMidEndSepPunct{\mcitedefaultmidpunct}
{\mcitedefaultendpunct}{\mcitedefaultseppunct}\relax
\EndOfBibitem
\bibitem[Shaw \emph{et~al.}(1967)Shaw, Reilly, Muysson, Pattenden, and Campbell]{shaw1967}
D.~M. Shaw, G.~A. Reilly, J.~R. Muysson, G.~E. Pattenden and F.~E. Campbell, \emph{Can. J. Earth Sci.}, 1967, \textbf{4}, 829--853\relax
\mciteBstWouldAddEndPuncttrue
\mciteSetBstMidEndSepPunct{\mcitedefaultmidpunct}
{\mcitedefaultendpunct}{\mcitedefaultseppunct}\relax
\EndOfBibitem
\bibitem[Shaw \emph{et~al.}(1976)Shaw, Dostal, and Keays]{shaw1976}
D.~M. Shaw, J.~Dostal and R.~R. Keays, \emph{GCA}, 1976, \textbf{40}, 73--83\relax
\mciteBstWouldAddEndPuncttrue
\mciteSetBstMidEndSepPunct{\mcitedefaultmidpunct}
{\mcitedefaultendpunct}{\mcitedefaultseppunct}\relax
\EndOfBibitem
\bibitem[McDonough and Sun(1995)]{mcdonough1995}
W.~F. McDonough and S.-s. Sun, \emph{Chem. Geol.}, 1995, \textbf{120}, 223--253\relax
\mciteBstWouldAddEndPuncttrue
\mciteSetBstMidEndSepPunct{\mcitedefaultmidpunct}
{\mcitedefaultendpunct}{\mcitedefaultseppunct}\relax
\EndOfBibitem
\bibitem[Wedepohl(1995)]{wedepohl1995}
K.~H. Wedepohl, \emph{GCA}, 1995, \textbf{59}, 1217--1232\relax
\mciteBstWouldAddEndPuncttrue
\mciteSetBstMidEndSepPunct{\mcitedefaultmidpunct}
{\mcitedefaultendpunct}{\mcitedefaultseppunct}\relax
\EndOfBibitem
\bibitem[Shallenberger(1996)]{shallenberger1996}
J.~R. Shallenberger, \emph{J. Vac. Sci. Technol. A}, 1996, \textbf{14}, 693--698\relax
\mciteBstWouldAddEndPuncttrue
\mciteSetBstMidEndSepPunct{\mcitedefaultmidpunct}
{\mcitedefaultendpunct}{\mcitedefaultseppunct}\relax
\EndOfBibitem
\bibitem[Verma \emph{et~al.}(2020)Verma, Kuwahara, Mori, Raja, and Yamashita]{verma2020}
P.~Verma, Y.~Kuwahara, K.~Mori, R.~Raja and H.~Yamashita, \emph{Nanoscale}, 2020, \textbf{12}, 11333\relax
\mciteBstWouldAddEndPuncttrue
\mciteSetBstMidEndSepPunct{\mcitedefaultmidpunct}
{\mcitedefaultendpunct}{\mcitedefaultseppunct}\relax
\EndOfBibitem
\bibitem[Khodakov \emph{et~al.}(2002)Khodakov, Griboval-Constant, Bechara, and Zholobenko]{khodakov2002}
A.~Y. Khodakov, A.~Griboval-Constant, R.~Bechara and V.~L. Zholobenko, \emph{J. Catal.}, 2002, \textbf{206}, 230--241\relax
\mciteBstWouldAddEndPuncttrue
\mciteSetBstMidEndSepPunct{\mcitedefaultmidpunct}
{\mcitedefaultendpunct}{\mcitedefaultseppunct}\relax
\EndOfBibitem
\bibitem[Liu \emph{et~al.}(2024)Liu, Wu, You, and Li]{liu2024}
Y.~Liu, F.~Wu, Z.~You and J.~Li, \emph{Catal. Lett.}, 2024, \textbf{154}, 5508--5520\relax
\mciteBstWouldAddEndPuncttrue
\mciteSetBstMidEndSepPunct{\mcitedefaultmidpunct}
{\mcitedefaultendpunct}{\mcitedefaultseppunct}\relax
\EndOfBibitem
\bibitem[P\'{e}rez-Estrada \emph{et~al.}(2024)P\'{e}rez-Estrada, Vargas-Villagr\'{a}n, Mendoza-Cruz, and Klimova]{perezestrada2024}
D.~E. P\'{e}rez-Estrada, H.~Vargas-Villagr\'{a}n, R.~Mendoza-Cruz and T.~E. Klimova, \emph{Nanoscale}, 2024, \textbf{16}, 11575--11591\relax
\mciteBstWouldAddEndPuncttrue
\mciteSetBstMidEndSepPunct{\mcitedefaultmidpunct}
{\mcitedefaultendpunct}{\mcitedefaultseppunct}\relax
\EndOfBibitem
\bibitem[Chandrashekhar \emph{et~al.}(2022)Chandrashekhar, Senthamarai, Kadam, Malina, Ka\v{s}l\'{i}k, Zbo\v{r}il, Gawande, Jagadeesh, and Beller]{chandrashekhar2022}
V.~G. Chandrashekhar, T.~Senthamarai, R.~G. Kadam, O.~Malina, J.~Ka\v{s}l\'{i}k, R.~Zbo\v{r}il, M.~B. Gawande, R.~V. Jagadeesh and M.~Beller, \emph{Nat. Catal.}, 2022, \textbf{5}, 20--29\relax
\mciteBstWouldAddEndPuncttrue
\mciteSetBstMidEndSepPunct{\mcitedefaultmidpunct}
{\mcitedefaultendpunct}{\mcitedefaultseppunct}\relax
\EndOfBibitem
\bibitem[Weissenrieder \emph{et~al.}(2005)Weissenrieder, Kaya, Lu, Gao, Shaikhutdinov, Freund, Sierka, Todorova, and Sauer]{weissenrieder2005}
J.~Weissenrieder, S.~Kaya, J.~L. Lu, H.~J. Gao, S.~Shaikhutdinov, H.~J. Freund, M.~Sierka, T.~K. Todorova and J.~Sauer, \emph{Phys. Rev. Lett.}, 2005, \textbf{95}, 076103\relax
\mciteBstWouldAddEndPuncttrue
\mciteSetBstMidEndSepPunct{\mcitedefaultmidpunct}
{\mcitedefaultendpunct}{\mcitedefaultseppunct}\relax
\EndOfBibitem
\bibitem[L{\"o}ffler \emph{et~al.}(2010)L{\"o}ffler, Uhlrich, Baron, Yang, Yu, Lichtenstein, Heinke, B{\"u}chner, Heyde, Shaikhutdinov,\emph{et~al.}]{loffler2010}
D.~L{\"o}ffler, J.~J. Uhlrich, M.~Baron, B.~Yang, X.~Yu, L.~Lichtenstein, L.~Heinke, C.~B{\"u}chner, M.~Heyde, S.~Shaikhutdinov \emph{et~al.}, \emph{{P}hys.~{R}ev.~{L}ett.}, 2010, \textbf{105}, 146104\relax
\mciteBstWouldAddEndPuncttrue
\mciteSetBstMidEndSepPunct{\mcitedefaultmidpunct}
{\mcitedefaultendpunct}{\mcitedefaultseppunct}\relax
\EndOfBibitem
\bibitem[Yang \emph{et~al.}(2012)Yang, Kaden, Yu, Boscoboinik, Martynova, Lichtenstein, Heyde, Sterrer, Włodarczyk, Sierka, Sauer, Shaikhutdinov, and Freund]{yang2012}
B.~Yang, W.~E. Kaden, X.~Yu, J.~A. Boscoboinik, Y.~Martynova, L.~Lichtenstein, M.~Heyde, M.~Sterrer, R.~Włodarczyk, M.~Sierka, J.~Sauer, S.~Shaikhutdinov and H.-J. Freund, \emph{Phys. Chem. Chem. Phys.}, 2012, \textbf{14}, 11344--11351\relax
\mciteBstWouldAddEndPuncttrue
\mciteSetBstMidEndSepPunct{\mcitedefaultmidpunct}
{\mcitedefaultendpunct}{\mcitedefaultseppunct}\relax
\EndOfBibitem
\bibitem[Altman \emph{et~al.}(2013)Altman, G\"{o}tzen, Samudrala, and Schwarz]{altman2013}
E.~I. Altman, J.~G\"{o}tzen, N.~Samudrala and U.~D. Schwarz, \emph{J. Phys. Chem. C}, 2013, \textbf{117}, 26144--26155\relax
\mciteBstWouldAddEndPuncttrue
\mciteSetBstMidEndSepPunct{\mcitedefaultmidpunct}
{\mcitedefaultendpunct}{\mcitedefaultseppunct}\relax
\EndOfBibitem
\bibitem[Yu \emph{et~al.}(2012)Yu, Yang, Boscoboinik, Shaikhutdinov, and Freund]{yu2012}
X.~Yu, B.~Yang, J.~A. Boscoboinik, S.~Shaikhutdinov and H.-J. Freund, \emph{Appl. Phys. Lett.}, 2012, \textbf{100}, 151608\relax
\mciteBstWouldAddEndPuncttrue
\mciteSetBstMidEndSepPunct{\mcitedefaultmidpunct}
{\mcitedefaultendpunct}{\mcitedefaultseppunct}\relax
\EndOfBibitem
\bibitem[Crampton \emph{et~al.}(2015)Crampton, Ridge, R\"{o}tzer, Zwaschka, Braun, D'Elia, Basset, Schweinberger, G\"{u}nther, and Heiz]{crampton2015}
A.~S. Crampton, C.~J. Ridge, M.~D. R\"{o}tzer, G.~Zwaschka, T.~Braun, V.~D'Elia, J.-M. Basset, F.~F. Schweinberger, S.~G\"{u}nther and U.~Heiz, \emph{J. Phys. Chem. C}, 2015, \textbf{119}, 13665--13669\relax
\mciteBstWouldAddEndPuncttrue
\mciteSetBstMidEndSepPunct{\mcitedefaultmidpunct}
{\mcitedefaultendpunct}{\mcitedefaultseppunct}\relax
\EndOfBibitem
\bibitem[Doudin \emph{et~al.}(2022)Doudin, Saritas, Li, Ennen, Boscoboinik, Dementyev, H\"{u}tten, Ismail-Beigi, and Altman]{doudin2022}
N.~Doudin, K.~Saritas, M.~Li, I.~Ennen, J.~A. Boscoboinik, P.~Dementyev, A.~H\"{u}tten, S.~Ismail-Beigi and E.~I. Altman, \emph{ACS mater. lett.}, 2022, \textbf{4}, 1660--1667\relax
\mciteBstWouldAddEndPuncttrue
\mciteSetBstMidEndSepPunct{\mcitedefaultmidpunct}
{\mcitedefaultendpunct}{\mcitedefaultseppunct}\relax
\EndOfBibitem
\bibitem[Jhang \emph{et~al.}(2017)Jhang, Zhou, Dagdeviren, Hutchings, Schwarz, and Altman]{jhang2017}
J.-H. Jhang, C.~Zhou, O.~E. Dagdeviren, G.~S. Hutchings, U.~D. Schwarz and E.~I. Altman, \emph{Phys. Chem. Chem. Phys}, 2017, \textbf{19}, 14001--14011\relax
\mciteBstWouldAddEndPuncttrue
\mciteSetBstMidEndSepPunct{\mcitedefaultmidpunct}
{\mcitedefaultendpunct}{\mcitedefaultseppunct}\relax
\EndOfBibitem
\bibitem[Tissot \emph{et~al.}(2018)Tissot, Weng, Schlexer, Pacchioni, Shaikhutdinov, and Freund]{tissot2018}
H.~Tissot, X.~Weng, P.~Schlexer, G.~Pacchioni, S.~Shaikhutdinov and H.-J. Freund, \emph{Surf. Sci.}, 2018, \textbf{678}, 118--123\relax
\mciteBstWouldAddEndPuncttrue
\mciteSetBstMidEndSepPunct{\mcitedefaultmidpunct}
{\mcitedefaultendpunct}{\mcitedefaultseppunct}\relax
\EndOfBibitem
\bibitem[Krinninger \emph{et~al.}(2024)Krinninger, Kraushofer, Refvik, Blum, and Lechner]{krinninger2024}
M.~Krinninger, F.~Kraushofer, N.~B. Refvik, M.~Blum and B.~A.~J. Lechner, \emph{ACS Appl. Mater. Interfaces}, 2024, \textbf{16}, 27481--27489\relax
\mciteBstWouldAddEndPuncttrue
\mciteSetBstMidEndSepPunct{\mcitedefaultmidpunct}
{\mcitedefaultendpunct}{\mcitedefaultseppunct}\relax
\EndOfBibitem
\bibitem[Włodarczyk \emph{et~al.}(2013)Włodarczyk, Sauer, Yu, Boscoboinik, Yang, Shaikhutdinov, and Freund]{wlodarczyk2013}
R.~Włodarczyk, J.~Sauer, X.~Yu, J.~A. Boscoboinik, B.~Yang, S.~Shaikhutdinov and H.-J. Freund, \emph{J. Am. Chem. Soc.}, 2013, \textbf{135}, 51\relax
\mciteBstWouldAddEndPuncttrue
\mciteSetBstMidEndSepPunct{\mcitedefaultmidpunct}
{\mcitedefaultendpunct}{\mcitedefaultseppunct}\relax
\EndOfBibitem
\bibitem[B\"{u}chner and Heyde(2017)]{buchner2017}
C.~B\"{u}chner and M.~Heyde, \emph{Prog. Surf. Sci.}, 2017, \textbf{92}, 341--374\relax
\mciteBstWouldAddEndPuncttrue
\mciteSetBstMidEndSepPunct{\mcitedefaultmidpunct}
{\mcitedefaultendpunct}{\mcitedefaultseppunct}\relax
\EndOfBibitem
\bibitem[Zhong and Freund(2022)]{zhong2022}
J.~Zhong and H.~Freund, \emph{Chem. Rev.}, 2022, \textbf{122}, 11172--11246\relax
\mciteBstWouldAddEndPuncttrue
\mciteSetBstMidEndSepPunct{\mcitedefaultmidpunct}
{\mcitedefaultendpunct}{\mcitedefaultseppunct}\relax
\EndOfBibitem
\bibitem[Altman and Dementyev(2024)]{altman2024}
E.~I. Altman and P.~Dementyev, \emph{Catal. Lett.}, 2024, \textbf{154}, 1359--1374\relax
\mciteBstWouldAddEndPuncttrue
\mciteSetBstMidEndSepPunct{\mcitedefaultmidpunct}
{\mcitedefaultendpunct}{\mcitedefaultseppunct}\relax
\EndOfBibitem
\bibitem[Huang \emph{et~al.}(2012)Huang, Kurasch, Srivastava, Skakalova, Kotakoski, Krasheninnikov, Hovden, Mao, Meyer, Smet, Muller, and Kaiser]{huang2012}
P.~Y. Huang, S.~Kurasch, A.~Srivastava, V.~Skakalova, J.~Kotakoski, A.~V. Krasheninnikov, R.~Hovden, Q.~Mao, J.~C. Meyer, J.~Smet, D.~A. Muller and U.~Kaiser, \emph{Nano Lett.}, 2012, \textbf{12}, 1081--1086\relax
\mciteBstWouldAddEndPuncttrue
\mciteSetBstMidEndSepPunct{\mcitedefaultmidpunct}
{\mcitedefaultendpunct}{\mcitedefaultseppunct}\relax
\EndOfBibitem
\bibitem[Jablonski and Powell(2010)]{jablonski2010}
A.~Jablonski and C.~Powell, \emph{National Institute of Standards and Technology, Gaithersburg}, 2010\relax
\mciteBstWouldAddEndPuncttrue
\mciteSetBstMidEndSepPunct{\mcitedefaultmidpunct}
{\mcitedefaultendpunct}{\mcitedefaultseppunct}\relax
\EndOfBibitem
\bibitem[Tanuma \emph{et~al.}(1991)Tanuma, Powell, and Penn]{tanuma1991}
S.~Tanuma, C.~J. Powell and D.~R. Penn, \emph{Surf. Interface Anal.}, 1991, \textbf{17}, 927--939\relax
\mciteBstWouldAddEndPuncttrue
\mciteSetBstMidEndSepPunct{\mcitedefaultmidpunct}
{\mcitedefaultendpunct}{\mcitedefaultseppunct}\relax
\EndOfBibitem
\bibitem[Tschersich and von Bonin(1998)]{tschersich1998}
K.~G. Tschersich and V.~von Bonin, \emph{J. Appl. Phys.}, 1998, \textbf{84}, 4065--4070\relax
\mciteBstWouldAddEndPuncttrue
\mciteSetBstMidEndSepPunct{\mcitedefaultmidpunct}
{\mcitedefaultendpunct}{\mcitedefaultseppunct}\relax
\EndOfBibitem
\bibitem[Tschersich(2000)]{tschersich2000}
K.~G. Tschersich, \emph{J. Appl. Phys.}, 2000, \textbf{87}, 2565--2573\relax
\mciteBstWouldAddEndPuncttrue
\mciteSetBstMidEndSepPunct{\mcitedefaultmidpunct}
{\mcitedefaultendpunct}{\mcitedefaultseppunct}\relax
\EndOfBibitem
\bibitem[Libra(2021)]{kolxpd}
J.~Libra, \emph{KolXPD Version 1.8.0 (Build 68)}, 2021\relax
\mciteBstWouldAddEndPuncttrue
\mciteSetBstMidEndSepPunct{\mcitedefaultmidpunct}
{\mcitedefaultendpunct}{\mcitedefaultseppunct}\relax
\EndOfBibitem
\bibitem[Ne\v{c}as and Klapetek(2012)]{gwyddion}
D.~Ne\v{c}as and P.~Klapetek, \emph{Open Phys.}, 2012, \textbf{10}, 181--188\relax
\mciteBstWouldAddEndPuncttrue
\mciteSetBstMidEndSepPunct{\mcitedefaultmidpunct}
{\mcitedefaultendpunct}{\mcitedefaultseppunct}\relax
\EndOfBibitem
\bibitem[Barinov \emph{et~al.}(2009)Barinov, Malcio\v{g}lu, Fabris, Sun, Gregoratti, Dalmiglio, and Kiskinova]{barinov2009}
A.~Barinov, O.~B. Malcio\v{g}lu, S.~Fabris, T.~Sun, L.~Gregoratti, M.~Dalmiglio and M.~Kiskinova, \emph{J. Phys. Chem. C}, 2009, \textbf{113}, 9009--9013\relax
\mciteBstWouldAddEndPuncttrue
\mciteSetBstMidEndSepPunct{\mcitedefaultmidpunct}
{\mcitedefaultendpunct}{\mcitedefaultseppunct}\relax
\EndOfBibitem
\bibitem[Vinogradov \emph{et~al.}(2011)Vinogradov, Schulte, Ng, Mikkelsen, Lundgren, Mårtensson, and Preobrajenski]{vinogradov2011}
N.~A. Vinogradov, K.~Schulte, M.~L. Ng, A.~Mikkelsen, E.~Lundgren, N.~Mårtensson and A.~B. Preobrajenski, \emph{J. Phys. Chem. C}, 2011, \textbf{115}, 9568--9577\relax
\mciteBstWouldAddEndPuncttrue
\mciteSetBstMidEndSepPunct{\mcitedefaultmidpunct}
{\mcitedefaultendpunct}{\mcitedefaultseppunct}\relax
\EndOfBibitem
\bibitem[Larciprete \emph{et~al.}(2012)Larciprete, Lacovig, Gardonio, Baraldi, and Lizzit]{larciprete2012}
R.~Larciprete, P.~Lacovig, S.~Gardonio, A.~Baraldi and S.~Lizzit, \emph{J. Phys. Chem. C}, 2012, \textbf{116}, 9900--9908\relax
\mciteBstWouldAddEndPuncttrue
\mciteSetBstMidEndSepPunct{\mcitedefaultmidpunct}
{\mcitedefaultendpunct}{\mcitedefaultseppunct}\relax
\EndOfBibitem
\bibitem[Finster \emph{et~al.}(1985)Finster, Schulze, Bechstedt, and Meisel]{finster1985}
J.~Finster, D.~Schulze, F.~Bechstedt and A.~Meisel, \emph{Surf. Sci.}, 1985, \textbf{152--153}, 1063--1070\relax
\mciteBstWouldAddEndPuncttrue
\mciteSetBstMidEndSepPunct{\mcitedefaultmidpunct}
{\mcitedefaultendpunct}{\mcitedefaultseppunct}\relax
\EndOfBibitem
\bibitem[Ni \emph{et~al.}(2012)Ni, Li, and Yang]{ni2012}
S.~Ni, Z.~Li and J.~Yang, \emph{Nanoscale}, 2012, \textbf{4}, 1184--1189\relax
\mciteBstWouldAddEndPuncttrue
\mciteSetBstMidEndSepPunct{\mcitedefaultmidpunct}
{\mcitedefaultendpunct}{\mcitedefaultseppunct}\relax
\EndOfBibitem
\bibitem[Nowakowski \emph{et~al.}(1992)Nowakowski, Vohs, and Bonnell]{nowakowski1992}
M.~Nowakowski, J.~Vohs and D.~Bonnell, \emph{Surf. Sci.}, 1992, \textbf{271}, L351--L356\relax
\mciteBstWouldAddEndPuncttrue
\mciteSetBstMidEndSepPunct{\mcitedefaultmidpunct}
{\mcitedefaultendpunct}{\mcitedefaultseppunct}\relax
\EndOfBibitem
\bibitem[Mohamed~Ibrahim \emph{et~al.}(2022)Mohamed~Ibrahim, Morisset, Baouche, and Dulieu]{mohamed2022}
A.~S. Mohamed~Ibrahim, S.~Morisset, S.~Baouche and F.~Dulieu, \emph{J. Chem. Phys.}, 2022, \textbf{156}, 194307\relax
\mciteBstWouldAddEndPuncttrue
\mciteSetBstMidEndSepPunct{\mcitedefaultmidpunct}
{\mcitedefaultendpunct}{\mcitedefaultseppunct}\relax
\EndOfBibitem
\bibitem[Johns \emph{et~al.}(2013)Johns, Alaboson, Patwardhan, Ryder, Schatz, and Hersam]{johns2013}
J.~E. Johns, J.~M.~P. Alaboson, S.~Patwardhan, C.~R. Ryder, G.~C. Schatz and M.~C. Hersam, \emph{J. Am. Chem. Soc.}, 2013, \textbf{135}, 18121--18125\relax
\mciteBstWouldAddEndPuncttrue
\mciteSetBstMidEndSepPunct{\mcitedefaultmidpunct}
{\mcitedefaultendpunct}{\mcitedefaultseppunct}\relax
\EndOfBibitem
\bibitem[Lander and Morrison(1962)]{lander1962}
J.~J. Lander and J.~Morrison, \emph{J. Appl. Phys.}, 1962, \textbf{33}, 2089--2092\relax
\mciteBstWouldAddEndPuncttrue
\mciteSetBstMidEndSepPunct{\mcitedefaultmidpunct}
{\mcitedefaultendpunct}{\mcitedefaultseppunct}\relax
\EndOfBibitem
\bibitem[Streit and Allen(1987)]{streit1987}
D.~C. Streit and F.~G. Allen, \emph{J. Appl. Phys.}, 1987, \textbf{61}, 2894--2897\relax
\mciteBstWouldAddEndPuncttrue
\mciteSetBstMidEndSepPunct{\mcitedefaultmidpunct}
{\mcitedefaultendpunct}{\mcitedefaultseppunct}\relax
\EndOfBibitem
\bibitem[Chaboy \emph{et~al.}(1995)Chaboy, Benfatto, and Davoli]{chaboy1995theoretical}
J.~Chaboy, M.~Benfatto and I.~Davoli, \emph{Phys. Rev. B}, 1995, \textbf{52}, 10014--10020\relax
\mciteBstWouldAddEndPuncttrue
\mciteSetBstMidEndSepPunct{\mcitedefaultmidpunct}
{\mcitedefaultendpunct}{\mcitedefaultseppunct}\relax
\EndOfBibitem
\bibitem[Li \emph{et~al.}(1993)Li, Bancroft, Kasrai, Fleet, Feng, Tan, and Yang]{li1993high}
D.~Li, G.~M. Bancroft, M.~Kasrai, M.~E. Fleet, X.~H. Feng, K.~H. Tan and B.~X. Yang, \emph{Solid State Commun.}, 1993, \textbf{87}, 613--617\relax
\mciteBstWouldAddEndPuncttrue
\mciteSetBstMidEndSepPunct{\mcitedefaultmidpunct}
{\mcitedefaultendpunct}{\mcitedefaultseppunct}\relax
\EndOfBibitem
\bibitem[Brown \emph{et~al.}(1977)Brown, Bachrach, and Skibowski]{Brown1977}
F.~C. Brown, R.~Z. Bachrach and M.~Skibowski, \emph{{P}hys.~{R}ev.~{B}}, 1977, \textbf{15}, 4781--4788\relax
\mciteBstWouldAddEndPuncttrue
\mciteSetBstMidEndSepPunct{\mcitedefaultmidpunct}
{\mcitedefaultendpunct}{\mcitedefaultseppunct}\relax
\EndOfBibitem
\bibitem[Garvie and Buseck(1999)]{garvie1999}
L.~A. Garvie and P.~R. Buseck, \emph{{A}m.~{M}in}, 1999, \textbf{84}, 946--964\relax
\mciteBstWouldAddEndPuncttrue
\mciteSetBstMidEndSepPunct{\mcitedefaultmidpunct}
{\mcitedefaultendpunct}{\mcitedefaultseppunct}\relax
\EndOfBibitem
\end{mcitethebibliography}
\bibliographystyle{rsc} 

\end{document}